\documentclass[a4paper,12pt]{article}
\pdfoutput=1
\usepackage{amssymb}
\usepackage{amsmath}
\usepackage{epsfig}
\usepackage{graphicx}
\usepackage{slashed}
\usepackage{xcolor}

\makeatletter

\@addtoreset{equation}{section}
\makeatother
\def\be{\begin{equation}}
\def\ee{\end{equation}}
\def\bea{\begin{eqnarray}}
\def\eea{\end{eqnarray}}

\def\({\left(}
\def\){\right)}
\def\<{\left<}
\def\>{\right>}

\def\tr{{\mbox{tr}}}
\def\be{\begin{equation}}
\def\ee{\end{equation}}
\def\bea{\begin{eqnarray*}}
\def\eea{\end{eqnarray*}}
\def\ben{\begin{eqnarray}}
\def\een{\end{eqnarray}}
\def\({\left(}
\def\){\right)}
\def\<{\left<}
\def\>{\right>}
\def\!{\right|}
\def\|{\left|}

\def\[{\left[}
\def\]{\right]}

\def\+{\bar}
\def\mb{\mathbb}
\def\tr{{\mbox{tr}}}

\def\L{{\cal{L}}}
\def\t{\widetilde}

\def\R{{\cal{R}}}

\def\N{{\cal{N}}}

\def\F{{\cal{F}}}

\def\H{{\cal{H}}}

\def\L{{\cal{L}}}

\def\E{{\cal{E}}}

\def\F{{\cal{F}}}
\def\Z{{\cal{Z}}}

\def\h{\widehat}

\def\ch{\mbox{ch}}

\def\a{\alpha}
\def\I{{\cal{I}}}
\def\Z{{\cal{Z}}}

\begin{document}

\setlength{\unitlength}{1mm}

\begin{titlepage}
\pagestyle{empty}
\vskip-10pt
\vskip-10pt
\hfill 
\begin{center}
\vskip 3truecm
{\Large \bf
Witten indices of abelian M5 brane on $\mb{R}\times S^5$}
\vskip 1cm
{\large \bf
Dongsu Bak,$^{\, \tt a,b}$ Andreas Gustavsson$^{\, \tt b}$}
\vskip0.8cm
\centerline{\sl  a) Physics Department,
University of Seoul, Seoul 02504 \rm KOREA}
 \vskip0.4cm
 \centerline{\sl b)
B.W. Lee Center for Fields, Gravity \& Strings}
 \centerline{\sl 
Institute for Basic Sciences, Daejeon 34047 \rm KOREA}
\vskip 0.1cm
\begin{center}
(\tt dsbak@uos.ac.kr, agbrev@gmail.com)
\end{center}
\end{center}
\vskip 2truecm
{\abstract{Witten indices and partition functions are computed for abelian 6d tensor and hypermultiplets on $\mb{R}\times S^5$ in Lorentzian signature in an R gauge field background which preserves some supersymmetry. We consider a general supersymmetric squashing that also admits squashing of the Hopf fiber. Wick rotation to Euclidean M5 brane amounts to Wick rotation of squashing parameters and the hypermultiplet mass parameter. We compute Casimir energies for tensor and hypermultiplets separately for general squashing, and match these with the corresponding gravitational anomaly polynomials. We extract Witten indices on $\mb{R}\times \mb{CP}^2$ and find that this is zero, again matching with the vanishing anomaly polynomial on an odd dimensional space.}}

\vfill
\vskip4pt
\end{titlepage}

\section{Introduction}
It has been conjectured that 6d $(2,0)$ theory \cite{Witten:1996hc} compactified on a circle, is equivalent to the dimensionally reduced 5d SYM theory \cite{Lambert:2010iw,Douglas:2010iu}. Confirming checks of this conjecture have been done for the abelian case in \cite{Kim:2011mv,Bak:2012ct,Kim:2013nva}. We understand the non-abelian 6d theory in the large $N$ limit by the AdS supergravity dual, so we can test the duality in the large $N$ limit. In \cite{Minahan:2013jwa} there was such an attempt. This paper took the large $N$ limit of the partition function of a round $S^5$ that was obtained in \cite{Kallen:2012cs,Kallen:2012va} for 5d SYM on round $S^5$ \cite{Hosomichi:2012ek}, and extracted the free energy. The partition function of squashed $S^5$ was obtained in \cite{Lockhart:2012vp,Kim:2012qf}. The corresponding 6d object is the superconformal index \cite{Bhattacharya:2008zy,Kim:2012ava,Kim:2012tr,Kim:2012qf,Kim:2013nva,Kim:2016usy}. For the round sphere case, an agreement with the corresponding supergravity computation in Euclidean AdS$_7$, could be found if the hypermultiplet mass parameter in the 5d SYM theory was Wick rotated as we go from Lorentzian M5 to Euclidean M5 theory \cite{Minahan:2013jwa}. But no justification for this Wick rotation from the 6d perspective was given. This result was our original motivation to better understand the 6d superconformal index in both Lorentzian and Euclidean signatures, as a continuation of \cite{Gustavsson:2015fra}.

\subsection{Radial quantization}
The abelian M5 brane superconformal index was first obtained in \cite{Bhattacharya:2008zy}, and has been generalized in \cite{Benvenuti:2016dcs}\footnote{We would like to thank S. Benvenuti for bringing this reference to our attention.}. The computation was done in Euclidean signature using radial quantization by summing up the contributions from all BPS letters. The bosonic part of the superconformal group $SO(1,7)\supset SO(1,1) \times SO(6)$ has Cartan generators $\Delta$ of $SO(1,1)$ which is the scaling dimension, and three Cartans $j_i$ of the rotation group $SO(6)$ of $S^5$. We have two Cartans $R_1$ and $R_2$ of $SO(5)$ R symmetry. The BPS equation is 
\ben
\Delta - j_1 - j_2 - j_3 + 2 \(R_1 + R_2\) &=& 0\label{BPS}
\een
We use the following index notation $Q_{j_1 j_2 j_3}^{R_1 R_2}$ for spinors such that $j_i = \pm$ refers to $j_i = \pm\frac{1}{2}$. 

{The bosonic fields are a selfdual two-form gauge potential $B_{MN}$ and five scalar fields $\phi^A$. The fermionic fields are four real chiral fermions. It turns out that the selfdual field strength $H_{MNP}$ does not saturate the BPS bound.

We define the singlet single-particle index as
\bea
\I^{singlet}_L(\beta,m,a_i) &:=& \tr (-1)^F e^{-\beta \(\Delta+\frac{1-3a}{2}\(R_1+R_2\)+m\(R_1-R_2\)+\sum_{i=1}^3 a_i j_i \)}
\eea
with $a:=(a_1+a_2+a_3)/3$. }The BPS letters associated to the singlet supercharge $Q^{--}_{---}$ are summarized in the Table \ref{singlet} below,
\begin{table}[ht]
\caption{Singlet case, with $a=\frac{1}{3}\(a_1+a_2+a_3\)$}\label{singlet}
\begin{tabular}{c|ccc|c}
letter & $\Delta$ & $(j_1,j_2,j_3)$ & $(R_1,R_2)$ & $(-1)^F e^{-\beta \(\Delta+\frac{1}{2}\(R_1+R_2\)+m\(R_1-R_2\)\)}$\\
&&&& $e^{-\beta\(\sum_{i=1}^3 a_i j_i - \frac{3a}{2}\(R_1+R_2\)\)}$\\
\hline
$\partial_1+i\partial_2$ & 1 & (1,0,0) & (0,0) & $e^{-\beta\(1+a_1\)}$\\
$\partial_3+i\partial_4$ & 1 & (0,1,0) & (0,0) & $e^{-\beta\(1+a_2\)}$\\
$\partial_5+i\partial_6$ & 1 & (0,0,1) & (0,0) & $e^{-\beta\(1+a_3\)}$\\
\hline
$\phi^1+i\phi^2$ & 2 & (0,0,0) & (-1,0) & $e^{-\beta\(\frac{3}{2}-m\)}e^{-\frac{3\beta a}{2}}$\\
$\phi^3+i\phi^4$ & 2 & (0,0,0) & (0,-1) & $e^{-\beta\(\frac{3}{2}+m\)}e^{-\frac{3\beta a}{2}}$\\
\hline
$\psi^{--}_{++-}$ & $\frac{5}{2}$ & $\(\frac{1}{2},\frac{1}{2},-\frac{1}{2}\)$ & $\(-\frac{1}{2},-\frac{1}{2}\)$ & $-e^{-\beta \(2-a_3\)}e^{-3 \beta a}$\\
$\psi^{--}_{+-+}$ & $\frac{5}{2}$ & $\(\frac{1}{2},-\frac{1}{2},\frac{1}{2}\)$ & $\(-\frac{1}{2},-\frac{1}{2}\)$ & $-e^{-\beta \(2-a_2\)}e^{-3 \beta a}$\\
$\psi^{--}_{-++}$ & $\frac{5}{2}$ & $\(-\frac{1}{2},\frac{1}{2},\frac{1}{2}\)$ & $\(-\frac{1}{2},-\frac{1}{2}\)$ & $-e^{-\beta \(2-a_1\)}e^{-3 \beta a}$\\
\hline
$\(\Gamma^M \nabla_M \psi\)^{--}_{+++}$ & $\frac{7}{2}$ & $\(\frac{1}{2},\frac{1}{2},\frac{1}{2}\)$ & $\(-\frac{1}{2},-\frac{1}{2}\)$ & $e^{-3\beta} e^{-3\beta a}$
\end{tabular}
\end{table}
We can apply any number of derivatives on any field (which leads to a geometric sum), and in doing so we have to subtract the letter index corresponding to the fermionic equation of motion (since this particular combination of derivatives acting on the fermion is zero just by that fermionic equation of motion, so it shall not be counted as a BPS state). We get the single particle index
\bea
\I^{singlet}_L(\beta,m,a_i) &=& \frac{e^{-\frac{3\beta}{2}}e^{-\frac{3\beta a}{2}}\(e^{\beta m} + e^{-\beta m}\)-e^{-2\beta}e^{-3\beta a}\(e^{\beta a_3}+e^{\beta a_2}+e^{\beta a_1}\)+e^{-3\beta}e^{-3\beta a}}{\(1-e^{-\beta\(1+a_1\)}\)\(1-e^{-\beta\(1+a_2\)}\)\(1-e^{-\beta\(1+a_3\)}\)}
\eea
If we pick $a_1=a_2=a_3=a$, then
\bea
\I^{singlet}_L(\beta,m,a) &=& \frac{e^{-\frac{3\beta}{2}}e^{-\frac{3\beta a}{2}}\(e^{\beta m} + e^{-\beta m}\)-3e^{-2\beta}e^{-2\beta a}+e^{-3\beta}e^{-3\beta a}}{\(1-e^{-\beta\(1+a\)}\)^3}
\eea
and this simplifies further at $m = \pm(\frac{1}{2}+\frac{a}{2})$, where we get
\bea
\I^{singlet}_L(\beta,a) &=& \frac{e^{-\beta(1+a)}}{1-e^{-\beta(1+a)}}
\eea

{ We define the triplet single-particle index as
\bea
\I^{triplet,I}_L(\beta,m,a_i) &:=& \tr (-1)^F e^{-\beta \(\Delta+\frac{1+a}{2}\(R_1+R_2\)+m\(R_1-R_2\)+\sum_{i=1}^3 a_i j_i\)}
\eea
with $a:=a_1-a_2+a_3$.} This preserves the triplet supercharge $Q^{--}_{+-+}$ and the corresponding BPS table is presented in Table \ref{triplet}.
\begin{table}[ht]
\caption{Triplet case, with $a=a_1-a_2+a_3$}\label{triplet}
\begin{tabular}{c|ccc|c}
letter & $\Delta$ & $(j_1,j_2,j_3)$ & $(R_1,R_2)$ & $(-1)^F e^{-\beta \(\Delta+\frac{1}{2}\(R_1+R_2\)+m\(R_1-R_2\)\)}$\\
&&&& $e^{-\beta\(\sum_{i=1}^3 a_i j_i + \frac{a}{2}\(R_1+R_2\)\)}$\\
\hline
$\partial_1+i\partial_2$ & 1 & (1,0,0) & (0,0) & $e^{-\beta\(1+a_1\)}$\\
$\partial_3+i\partial_4$ & 1 & (0,1,0) & (0,0) & $e^{-\beta\(1+a_2\)}$\\
$\partial_5+i\partial_6$ & 1 & (0,0,1) & (0,0) & $e^{-\beta\(1+a_3\)}$\\
\hline
$\phi^1+i\phi^2$ & 2 & (0,0,0) & (-1,0) & $e^{-\beta\(\frac{3}{2}-m\)}e^{\frac{\beta a}{2}}$\\
$\phi^3+i\phi^4$ & 2 & (0,0,0) & (0,-1) & $e^{-\beta\(\frac{3}{2}+m\)}e^{\frac{\beta a}{2}}$\\
\hline
$\psi^{--}_{++-}$ & $\frac{5}{2}$ & $\(\frac{1}{2},\frac{1}{2},-\frac{1}{2}\)$ & $\(-\frac{1}{2},-\frac{1}{2}\)$ & $-e^{-2\beta}e^{\beta \(a_3-a_2\)}$\\
$\psi^{--}_{+-+}$ & $\frac{5}{2}$ & $\(\frac{1}{2},-\frac{1}{2},\frac{1}{2}\)$ & $\(-\frac{1}{2},-\frac{1}{2}\)$ & $-e^{-2\beta}$\\
$\psi^{--}_{-++}$ & $\frac{5}{2}$ & $\(-\frac{1}{2},\frac{1}{2},\frac{1}{2}\)$ & $\(-\frac{1}{2},-\frac{1}{2}\)$ & $-e^{-2\beta}e^{\beta \(a_1-a_2\)}$\\
\hline
$\(\Gamma^M \nabla_M \psi\)^{--}_{+++}$ & $\frac{7}{2}$ & $\(\frac{1}{2},\frac{1}{2},\frac{1}{2}\)$ & $\(-\frac{1}{2},-\frac{1}{2}\)$ & $e^{-3\beta} e^{-\beta a_2}$
\end{tabular}
\end{table}
From this table, we read off the full triplet letter index as 
\bea
\I^{triplet,I}_L(\beta,m,a_i) &=& \frac{e^{-\frac{3\beta}{2}}e^{\frac{\beta a 
}{2}}\(e^{\beta m} + e^{-\beta m}\)-e^{-2\beta}\(1+e^{\beta\(a_3-a_2\)}+e^{\beta \(a_1-a_2\)}\)+e^{-3\beta}e^{-\beta a_2}}{\(1-e^{-\beta\(1+a_1\)}\)\(1-e^{-\beta\(1+a_2\)}\)\(1-e^{-\beta\(1+a_3\)}\)}
\eea

Translation along the Hopf fiber of $S^5$ is generated by $j_1+j_2+j_3$ and corresponds to taking $a_1=a_2=a_3=a$. We may also consider another Hopf fibration of $S^5$. We may assign the generator $j_1-j_2-j_3$ as the generator for translations along this new Hopf fiber.  This new generator is obtained from the old generator by flipping the signs of $j_2$ and $j_3$. After this sign flip we get the supercharge $Q^{--}_{++-}$. To preserve it, we also need to modify the definition for $a$ to read $a := a_1+a_2-a_3$. The index with respect to the new Hopf fibration reads
\bea
\I^{triplet,II}_L(\beta,m,a_i) &=& \frac{e^{-\frac{3\beta}{2}}e^{\frac{\beta a
}{2}}\(e^{\beta m} + e^{-\beta m}\)-e^{-2\beta}\(1+e^{\beta\(a_2-a_3\)}+e^{\beta \(a_1+a_2\)}\)+e^{-3\beta}e^{+\beta a_2}}{\(1-e^{-\beta\(1+a_1\)}\)\(1-e^{-\beta\(1-a_2\)}\)\(1-e^{-\beta\(1-a_3\)}\)}
\eea

There is a third Hopf fibration which corresponds to taking its generator as $-j_1-j_2+j_3$ and which preserves the supercharge $Q_{-++}^{--}$. For this case we need to define $a:=-a_1+a_2+a_3$. The index with respect to this third Hopf fibration is given by 
\bea
\I^{triplet,III}_L(\beta,m,a_1,a_2,a_3) &=& f^{triplet,II}_L(\beta,m,a_3,a_2,a_1)
\eea
and it will be of the same form as $\I^{triplet,II}(\beta,m,a_1,a_2,a_3)$. Let us now evaluate these three indices for squashing along their respective Hopf fibers, at $a_1=a_2=a_3=a$. We find 
\bea
\I^{triplet,I}_L(\beta,a) &=& \frac{e^{-\frac{3\beta}{2}}e^{\frac{\beta a}{2}}\(e^{\beta m} + e^{-\beta m}\)-3e^{-2\beta}+e^{-3\beta}e^{-\beta a}}{\(1-e^{-\beta\(1+a\)}\)^3}\cr
\I^{triplet,II}_L(\beta,a) &=& \frac{e^{-\frac{3\beta}{2}}e^{\frac{\beta a}{2}}\(e^{\beta m} + e^{-\beta m}\)-e^{-2\beta}\(2+e^{2\beta a}\)+e^{-3\beta}e^{\beta a}}{\(1-e^{-\beta\(1-a\)}\)^2\(1-e^{-\beta\(1+a\)}\)}\cr
\I^{triplet,III}_L(\beta,a) &=& \I^{triplet,II}_L(\beta,a) 
\eea
We also notice the following relation between these indices,
\bea
\I^{singlet}(\beta,m,a,a,a) &=& \I^{triplet,I}(\beta,m,a,a,a)\cr
\I^{singlet}(\beta,m,-a,-a,a) &=& \I^{triplet,II}(\beta,m,a,a,a)
\eea
We have two points at $m = \pm(\frac{1}{2}+\frac{a}{2})$ where we have the simplification 
\bea
\I^{triplet,II}_L(\beta,a) &=& \frac{e^{-\beta(1-a)}}{1-e^{-\beta(1-a)}}
\eea

By a direct computation, we will reproduce these indices. However, there are a couple of surprises.

First, we obtain these indices with real-valued chemical potentials $m$ and $a_i$ only in Lorentzian signature where $\beta = iT$ with $T$ a real time interval. 

Second, when we compute the indices
\bea
\I^{singlet} &=& \tr (-1)^F e^{-\beta H} e^{-\beta a\(j-\frac{3}{2} \(R_1+R_2\)\)}\cr\I^{triplet} &=& \tr (-1)^F e^{-\beta H} e^{-\beta a \(j+\frac{1}{2} \(R_1+R_2\)\)}
\eea
with $j:=j_1+j_2+j_3$, and $H=\Delta+\frac{1}{2}\(R_1+R_2\)+m\(R_1-R_2\)$ (to be specified below), then we should expect to get results corresponding to $\I^{singlet}$ and $\I^{triplet,I}$ above. This is because we pick the same generator $j = j_1+j_2+j_3$ along the same Hopf fiber for the computation of both these indices. What explicit computations shows, is that while we do reproduce $\I^{singlet}$, instead of getting $\I^{triplet,I}$, we get $\I^{triplet,II}$ that from the viewpoint of radial quantization as presented above appears to correspond to a different Hopf fiber! This result is hard to understand from radial quantization alone. It shows limitations of radial quantization when applied to the M5 brane. We attribute these limitations of radial quantization to the fact that for the M5 brane there are no real-valued fields and no real-valued Lagrangian in Euclidean signature and this motivates us to do a direct computation in Lorentzian signature. 

\subsection{Wick rotation}
In Lorentzian signature we have a typical chemical potential of the form $e^{-iTa_1}$. Wick rotation is done by taking $\beta$ real and positive. Let us assume that $T$ is also positive. Then $T=\beta$ where left-hand side is the quantity in Lorentzian signature and the right-hand side is the same quantity in Euclidean signature. Then we have $e^{-\beta a^E_1} = e^{-T a^E_1}$ in Euclidean signature. We want this to be unchanged by Wick rotation. This amounts to taking 
\bea
a^E_1 &=& ia_1
\eea
The same goes through for all the chemical potentials. Thus $m^E = i m$. Thus in Euclidean signature the indices are given by
\bea
\I^{singlet}_E(\beta,m,a_i) &=& \frac{e^{-\frac{3\beta}{2}}e^{-\frac{3i\beta a}{2}}\(e^{i\beta m} + e^{-i\beta m}\)-e^{-2\beta}e^{-3i\beta a}\(e^{i\beta a_3}+e^{i\beta a_2}+e^{i\beta a_1}\)+e^{-3\beta}e^{-3i\beta a}}{\(1-e^{-\beta\(1+ia_1\)}\)\(1-e^{-\beta\(1+ia_2\)}\)\(1-e^{-\beta\(1+ia_3\)}\)}\cr
\I^{triplet,II}_E(\beta,m,a_i) &=& \frac{e^{-\frac{3\beta}{2}}e^{\frac{i\beta a}{2}}\(e^{i\beta m} + e^{-i\beta m}\)-e^{-2\beta}\(1+e^{i\beta\(a_2-a_3\)}+e^{i\beta \(a_1+a_2\)}\)+e^{-3\beta}e^{i\beta a_2}}{\(1-e^{-\beta\(1+ia_1\)}\)\(1-e^{-\beta\(1-ia_2\)}\)\(1-e^{-\beta\(1-ia_3\)}\)}
\eea
Moreover, 
\bea
\I^{singlet}_E(\beta,a) &=& \frac{e^{-\beta(1+ia)}}{1-e^{-\beta(1+ia)}}\cr
\I^{triplet,II}_E(\beta,a) &=& \frac{e^{-\beta(1-ia)}}{1-e^{-\beta(1-ia)}}
\eea
which both happen at the critical masses $\pm m = -\frac{i}{2}+\frac{a}{2}$. If we introduce a modular parameter $\tau = -\beta a + i \beta$, then
\bea
\I^{singlet}_E(\beta,a) &=& \frac{e^{i\tau}}{1-e^{i\tau}}
\eea
and $\I^{triplet,II}_E(\beta,a)$ is the complex conjugate of this. The full index can be extracted from the single particle index as follows
\bea
I(\beta,a) &=& e^{-\beta E_{M5}} PE[\I](\beta,a)
\eea
where 
\bea
PE[\I](\beta,a) &:=& \exp\[\sum_{n=1}^{\infty} \frac{1}{n} \I(\beta n,a)\]
\eea
is the plethystic exponent, and 
\bea
E_{M5} &=& -\frac{1+ia}{24}
\eea
is the Casimir energy. Explicitly, we get 
\bea
I^{singlet}_E(\beta,a) &=& e^{-\frac{i\tau}{24}} \prod_{n=1}^{\infty} \(1-e^{in\tau}\)^{-1} = \frac{1}{\eta(\tau)}
\eea
and $I^{triplet,II}$ is the complex conjugate of this. Although $E_{M5}$ is complex in Euclidean signature, it is real in Lorentzian signature and is given by equation (\ref{CasimirEnergy}). We thus see that we have got a familiar modular form on a two-torus. This two torus is spanned by the time axis and the Hopf circle, and $a$ is the translation along the Hopf circle that together with the Euclidean time axis makes up a slanted Euclidean torus. This modular form for the special case $a=0$ (that is, a torus with no slanting) was first obtained in \cite{Kim:2012ava} for the singlet case. What is new here is that we generalize this to a slanted two-torus and also include the triplet case.

Let us also note that the index for 6d $(0,2)$ theory is the complex conjugate of the index for the 6d $(2,0)$ theory; changing sign of $a$ corresponds to changing 6d chirality, as we will see in more detail later on.

\subsection{Hamiltonian computation}
In this paper we will not make use of the superconformal symmetry\footnote{There is no conformal map from Lorentzian $\mb{R}\times S^5$ to $\mb{R}^{1,5}$ so the conformal symmetry does not seem to be so helpful in Lorentzian signature.}. That is why we refer to the corresponding quantity as the Witten index, which we define as
\bea
\I &=& \tr (-1)^F e^{-i T H}
\eea
where $\tr$ is over single-particle states, and where we have the M5 brane Hamiltonian on $S^5$ that is evolving in Lorentzian time by an interval $T>0$. The space $S^5$ is compact and the spectra of the Laplace operators and of the Dirac operator on $S^5$ are discrete. The spectrum of the Hamiltonian is therefore also discrete and has a mass gap. From the commutator of two supercharges one may deduce a BPS equation. Since spectrum is discrete and we have supersymmetric pairing of non-BPS states, the Witten index picks up contributions only from those BPS states. 
 
If we consider round $S^5$, then in our R-gauge field background { (\ref{R gauge field})} we find that our Hamiltonian can be expressed in the form
\bea
H &=& H_0 + \frac{1}{2r} \(R_{12}+R_{34}\) + m\(R_{12}-R_{34}\)
\eea
where $H_0$ is the nonsupersymmetric Hamiltonian we compute with R gauge field turned off (and which would correspond to $\Delta$ in radial quantization). Here by $R_{12}$ and $R_{34}$ we denote the two Cartan generators of $SO(5)$ which is the R-symmetry of the M5 brane (and which we denoted as $R_1$ and $R_2$ above). However, a generic mass parameter $m$ breaks this symmetry down to $SU(2)_R\times SU(2)_F$ where $SU(2)_R$ is the R symmetry and $SU(2)_F$ is another global symmetry. The $8$ real supercharges ($\N=(1,0)$ supersymmetry) that we preserve do not commute with $H_0$. The purpose with turning on some R gauge field background is to preserve some supersymmetry. Thus supercharges that we preserve commute with $H$, and this is true for any value of the mass parameter $m$, so in particular they commute with $R_{12}-R_{34}$. The points at $\pm m=\frac{1}{2}$ have enhanced supersymmetry with $16$ real supercharges. On a squashed $S^5$ where the squashing is along the Hopf fiber, it is plausible that this critical value gets changed to the value $\pm m = \frac{1}{2} + \frac{a}{2}$ where we saw the simplification of the index was happening above. But to show this we would need to derive the Killing spinor solution on this squashed $S^5$. 
 
We can consider a general squashing on $S^5$ by turning on three chemical potentials $a_i = (a,b,c)$ for the three Cartan generators of the isometry group $SO(6)$, without imposing any restrictions on these chemical potentials other than they shall be real-valued. If we use a parametrization 
\bea
z_i &=& r_i e^{i\phi_i}
\eea
of $S^5$ with $r_i$ lying on $S^2$, say $r_i = r \(\sin\theta \sin \phi,\sin\theta \cos \phi,\cos\theta\)$, we get the induced metric 
\bea
ds^2 &=& dr_1^2 + r_1^2 d\phi_1^2 + dr_2^2 + r_2^2 d\phi_2^2 + dr_3^2 + r_3^2 d\phi_3^2
\eea
on $S^5$. In Lorentzian signature squashing by turning on these chemical potentials amounts to changing the metric on $\mb{R}\times S^5$ into \cite{Kim:2012qf}
\bea
ds^2 &=& - dt^2 + dr_1^2 + r_1^2 \(d\phi_1+a dt\)^2 + dr_2^2 + r_2^2 \(d\phi_2+b dt\)^2 + dr_3^2 + r_3^2 \(d\phi_3+c dt\)^2
\eea
Wick rotation of $t$ alone would lead to a complex metric, so we propose that we should Wick rotate the $a_i$'s as well so that the Euclidean metric becomes
\bea
ds^2 &=& dt^2 + dr_1^2 + r_1^2 \(d\phi_1+a dt\)^2 + dr_2^2 + r_2^2 \(d\phi_2+b dt\)^2 + dr_3^2 + r_3^2 \(d\phi_3+c dt\)^2
\eea
We have not been able to rigorously compute the Witten index with generic squashing parameters $a_i$ since the representation theory of $SO(6)$ is complicated. Instead we will compute the Witten index in Lorentzian signature with $a=b=c$ by compensating this with a certain R rotation that also depends on the parameter $a$, such that the whole rotation generator including the R  rotation, commutes with some amount of supersymmety. By combining this result with the result we obtained above using radial quantization for general squashing, it easy to make a guess the result with general squashing.

\section{The M5 brane Lagrangian}
The Abelian M5 brane on $\mb{R}\times S^5$ where $r$ is the radius of $S^5$, may be described by the following supersymmetric Lagrangian\footnote{{ In this Lagrangian we have included a decoupled wrong chirality part to the selfdual tensor gauge field as a spectator field.}}
\bea
\L &=& -\frac{1}{24} H_{MNP}^2 - \frac{1}{2} (D_M\phi^A)^2 + \frac{i}{2}\bar\psi\Gamma^M D_M\psi - \frac{2}{r^2}(\phi^A)^2
\eea
We use 11d gamma matrices where $M=(0,m)$ and $m=1,...,5$ is vector index on $S^5$ and $A=1,...,5$ is a vector index of the $SO(5)$ R-symmetry. The Dirac conjugate is $\bar\psi := \psi^{\dag}\Gamma^0$. We define the covariant derivative as $D_M = \nabla_M - iA_M$ where $A_M = \frac{1}{2}A_M^{AB} M^{AB}$ is the R-gauge field, $M^{AB}$ are $SO(5)$ generators\footnote{We normalize these generators so that $M^{AB}_{CD} = 2i \delta^{AB}_{CD}$ in the vector representation. Then we have $M^{AB} = \frac{i}{2}\h\Gamma^{AB}$ in the spinor representation.}, and we have the $(2,0)$ supersymmetry
\bea
\delta B_{MN} &=& i\bar\epsilon\Gamma_{MN}\psi\cr
\delta \phi^A &=& i\bar\epsilon\h\Gamma^A\psi\cr
\delta \psi &=& \frac{1}{12}\Gamma^{MNP}\epsilon H_{MNP}+\Gamma^M\h\Gamma^A\epsilon D_M \phi^A - \frac{2}{3} \h\Gamma^A (\Gamma^M D_M\epsilon)\phi^A
\eea
where 
\bea
D_M \epsilon &=& \frac{1}{6}\Gamma_M \Gamma^N D_N \epsilon\cr
\Gamma^{0\h 1\h 2\h 3\h 4\h 5} \epsilon &=& -\epsilon
\eea
where $\h m=\h 1,...,\h 5$ denote tangent space indices of $S^5$. We break the supersymmetry down to $(1,0)$ by the Weyl projections
\bea
\h\Gamma^{1234} \epsilon &=& -\epsilon\cr
\h\Gamma^{1234} \psi_{\pm} &=& \pm \psi_{\pm}
\eea
The $(2,0)$ tensor multiplet separates into one $(1,0)$ tensor multiplet
\bea
\L_{tensor} &=& -\frac{1}{24} H_{MNP}^2 - \frac{1}{2} (D_M\phi^5)^2 + \frac{i}{2}\bar\psi_-\Gamma^M D_M\psi_- - \frac{2}{r^2}(\phi^5)^2
\eea
\bea
\delta B_{MN} &=& i\bar\epsilon\Gamma_{MN}\psi_-\cr
\delta \phi^5 &=& i\bar\epsilon\h\Gamma^5\psi_-\cr
\delta \psi_- &=& \frac{1}{12}\Gamma^{MNP}\epsilon H_{MNP}+\Gamma^M\h\Gamma^5\epsilon D_M \phi^5 - \frac{2}{3} \h\Gamma^5 (\Gamma^M D_M\epsilon)\phi^5
\eea
and one $(1,0)$ hypermultiplet
\bea
\L_{hyper} &=& - \frac{1}{2} (D_M\phi^I)^2 + \frac{i}{2}\bar\psi_+\Gamma^M D_M\psi_+ - \frac{2}{r^2}(\phi^I)^2
\eea
\bea
\delta \phi^I &=& i\bar\epsilon\h\Gamma^I\psi_+\cr
\delta \psi_+ &=& \Gamma^M\h\Gamma^I\epsilon D_M \phi^I - \frac{2}{3} \h\Gamma^I (\Gamma^M D_M\epsilon)\phi^I
\eea
where $I=1,2,3,4$.

To go further, we need to pick an R gauge field and fix a gamma matrix convention. We pick the nonzero R gauge field components as
\ben
A_0^{12} &=& \frac{1}{2r}+m\cr
A_0^{34} &=& \frac{1}{2r}-m\label{R gauge field}
\een
and fix the gamma matrices in terms of Pauli sigma matrices $(\sigma^1,\sigma^2,\sigma^3)$ as
\bea
\Gamma^0 &=& i\sigma^2 \otimes 1 \otimes 1\cr
\Gamma^m &=& \sigma^1 \otimes \gamma^m \otimes 1
\eea
so that 
\bea
\Gamma^{012345} &=& \sigma^3 \otimes 1\otimes 1
\eea
and we choose
\bea
\h\Gamma^A &=& \sigma^3 \otimes 1 \otimes \h\gamma^A
\eea
so that
\bea
\h\Gamma^{1234} &=& 1\otimes 1\otimes \h\gamma^{1234}
\eea
Finally we choose
\bea
\h\gamma^I &=& \(\begin{array}{cc}
0 & \sigma^I\\
(\sigma^I)^{\dag} & 0
\end{array}\)
\eea
where $\sigma^I := (\sigma^1,\sigma^2,\sigma^3,-i)$.

With this setup, we get
\bea
\L_{tensor} &=& -\frac{1}{24}H_{MNP}^2-\frac{1}{2}(\nabla_M\phi^5)^2-\frac{2}{r^2}(\phi^5)^2\cr
&&-\frac{i}{2}\psi_-^{\dag}\dot\psi_-+\frac{1}{4r}\psi_-^{\dag}\sigma^3\psi_-+\frac{i}{2}\psi_-^{\dag}\gamma^m\nabla_m\psi_-
\eea
and with $I=(a,i)$, $a=1,2$, $i=3,4$, $\epsilon^{12}=\epsilon^{34}=1$,
\bea
\L_{hyper} &=& \frac{1}{2}(\dot\phi^a)^2 - \frac{1}{2} (\nabla_m\phi^a)^2 + \(\frac{1}{2r}+m\)\epsilon^{ab}\dot\phi^a \phi^b - \frac{1}{2}\(\frac{15}{4r^2}-\frac{m}{r}-m^2\)(\phi^a)^2\cr
&+& \frac{1}{2}(\dot\phi^i)^2 - \frac{1}{2} (\nabla_m\phi^i)^2 + \(\frac{1}{2r}-m\)\epsilon^{ij}\dot\phi^i \phi^j - \frac{1}{2}\(\frac{15}{4r^2}+\frac{m}{r}-m^2\)(\phi^i)^2\cr
&-&\frac{i}{2}\psi_+^{\dag}\dot\psi_++\frac{m}{2}\psi_+^{\dag}\sigma^3\psi_++\frac{i}{2}\psi_+^{\dag}\gamma^m\nabla_m\psi_+
\eea
We will now compute the contributions to the Witten index or the partition function for each field separately.

\section{Computation of the Witten index}
\subsection{The tensor multiplet scalar field contribution}
The Lagrangian for the tensor multiplet scalar field $\phi := \phi^5$ is given by 
\bea
\L &=& \frac{1}{2}\dot\phi^2 - \frac{1}{2} g^{mn} \partial_m\phi \partial_n \phi - \frac{2}{r^2} \phi^2
\eea
where $g_{mn}$ denotes the metric tensor on $S^5$ with radius $r$. We expand the scalar field in spherical scalar harmonics $Y_{n,m_1,m_2,m_3}$,
\bea
\phi &=& x + \sum_{n=1}^{\infty} \sum_{m_1,m_2,m_3} x_{n,m_1,m_2,m_3} Y_{n,m_1,m_2,m_3}
\eea
where we define 
\bea
x_{n,-m_1,-m_2,-m_3} &=& (x_{n,m_1,m_2,m_3})^*
\eea
ensuring that $\phi$ is a real-valued field. Here $m_i$ denote the $j_i$ charges that are the Cartans of $SO(6)$ isometry group of $S^5$. When we do the separation into $x_{n,m_1,m_2,m_3}$ and $(x_{n,m_1,m_2,m_3})^*$ we must cut the sum over $\vec{m} = (m_1,m_2,m_3)$ by half to avoid double counting of modes. Effectively this means that the degeneracy $b_n=\frac{1}{12}(n+1)(n+2)^2(n+3)$ (for how to get this dimension, see (\ref{degs}) in the appendix) of $Y_{n,m_1,m_2,m_3}$ is cut by half, to $\frac{b_n}{2}$. Now this number $b_n$ turns out to not always be an integer, for example $b_4 = 105$. Nevertheless it works, which we will explain in a moment. 

By noting the eigenvalues of the scalar Laplacian\footnote{Our definition of the Laplacian is $\triangle = dd^{\dag}+d^{\dag}d$. The scalar Laplacian is obtained by acting on a zero form and takes the form $\triangle = -g^{mn}D_mD_n$.} as given by eq (\ref{scalarlaplacian}) in the appendix, we find the Hamiltonian is composed of a zero mode part
\bea
H_{zero} &=& \frac{1}{2} p^2 + \frac{2}{r^2} x^2
\eea
and an oscillator mode part
\bea
H_{osc} &=& \sum_{n>0} \( (p_n)^* p_n + \(\frac{n+2}{r}\)^2 x_n (x_n)^*\)
\eea
Let us isolate one typical oscillator mode for which we have the Hamiltonian on the form
\bea
H &=& p^* p + \Omega^2 x x^*
\eea
We define 
\bea
a &=& \frac{1}{\sqrt{2\Omega}} \(p-i\Omega x^*\)\cr
b &=& \frac{1}{\sqrt{2\Omega}} \(p^* - i\Omega x\)
\eea
Then 
\bea
[a,a^*] &=& 1\cr
[b,b^*] &=& 1
\eea
and 
\bea
H &=& \Omega \(b^* b + a^* a + 1\)
\eea
Thus 
\bea
H_{zero} &=& \frac{2}{r} \(N+\frac{1}{2}\)\cr
H_{osc} &=& \sum_{n>0} \frac{n+2}{r} \(N_n^a + N_n^b + 1\)
\eea
where we define
\bea
N_n^a &=& a^{\dag}_n a_n\cr
N_n^b &=& b^{\dag}_n b_n
\eea
For the zero mode part there is just one set of creation and annihilation operators and $N$ denotes their corresponding number operator. 

{ We define the single particle index as 
\bea
\tr (-1)^F e^{-\beta \Delta H}
\eea
where $\tr$ is a sum over all states with $\sum N_n = 1$ for one-particle excitations, and with $\Delta H = H(N_n) - H(0)$ is the Hamiltonian of the single-particle excitation minus the Hamiltonian of the vacuum. This in particular means that the normal ordering constant does not enter in the single-particle index. We then compute the full index by taking the plethystic exponent, and multiplying this by an exponent of the Casimir energy as a prefactor. This prefactor contains the contributions of the zero point energies.}

The single particle index is given by\footnote{Here $r$ has been absorbed into $\beta$, or more simply, we put $r=1$.}
\bea
b_0 e^{-\beta 2} + \sum_{n>0} 2 \frac{b_n}{2} e^{-\beta\(n+2\)}
\eea
where the factor $\frac{b_n}{2}$ has argued for above. Here we also see explicitly that real modes can be viewed as $1/2$ of a complex mode. Namely the real zero mode, naturally gets incorporated in complex oscillator mode sum as we can write the sum in the following neat form
\bea
\sum_{n=0}^{\infty} b_n e^{-\beta\(n+2\)}
\eea
treating the zero mode and the complex oscillator modes and real oscillator modes on the same footing all entering in the same summation over $n$ here. This is the explanation why we can divide $b_n$ by $2$ even if $b_n$ is an odd integer. The reason is that when $b_n$ is odd, we have that $b_{n,m=0}$ is odd where $m=m_1+m_2+m_3$ (for example $b_n=105$ and $b_{n,0}=27$ and this pattern is easily seen to be general since $b_n = b_{n,0} + 2\sum_{m>0} b_{n,m}$. Thus if $b_n$ is odd, then so is $b_{n,0}$, for more details, see appendix A). But for those modes we can always find a rearrangement such that these modes are all real modes of degeneracy $b_{n,0}$ (rather than complex modes of degeneracy $\frac{b_{n,0}}{2}$, which would be problematic if this degeneracy is not an integer number).

\subsection{The tensor multiplet fermion contribution}
The Lagrangian is 
\bea
\L &=& -\frac{i}{2} \psi^{\dag} \dot\psi + \frac{i}{2} \psi^{\dag} \Gamma^{0m} \nabla_m \psi - \frac{i}{4r} \psi^{\dag} \h\Gamma^{12} \psi\cr
&=& - \frac{i}{2} (\psi^{\alpha}_a)^* \dot\psi^{\alpha}_a + \frac{i}{2} (\psi^{\alpha}_a)^* (\gamma^m)^{\alpha}{}_{\beta} \nabla_m \psi^{\beta}_a - \frac{i}{4r} (\psi^{\alpha}_a)^* (i\sigma^3)_a{}^b \psi^{\alpha}_b
\eea
We have the Majorana condition
\bea
(\psi^{\alpha}_a)^* &=& \psi^{\beta}_b C_{\beta\alpha} \epsilon^{ba}
\eea
We can use that to eliminate $\psi_-$ and replace it as follows,
\bea
(\psi_+^{\beta})^* &=& - \psi_-^{\beta} C_{\beta\alpha}
\eea
if we assume that $\epsilon^{+-}=1$. We then define a single component complex spinor 
\bea
\psi^\alpha &:=& \psi^\alpha_+
\eea
and then
\bea
\psi_- &=& - \psi^*
\eea
In terms of this single complex spinor, the Lagrangian becomes
\bea
\L &=& - i \psi^* \dot\psi + \psi^* i \gamma^m \nabla_m \psi - \frac{1}{2r} \psi^* \psi
\eea
We expand in spinor harmonics and get the Lagrangian for the modes as
\bea
L &=& \sum_n \(-i \psi_n^* \dot\psi_n - \(\mu_n + \frac{1}{2r}\) \psi_n^* \psi_n\)
\eea
Here $\mu_n$ denotes the eigenvalue of the Dirac operator acting on a spinor harmonic. This is given by \cite{Trautman:1995fr,Kim:2012ava} 
\bea
\mu_n &=& \pm\frac{1}{r}\(n + \frac{5}{2}\)
\eea
each sign comes with the degeneracy $f_n = \frac{1}{6}(n+1)(n+2)(n+3)(n+4)$. For more details on the spinor harmonics, we refer to the appendix and eq (\ref{degs}) there. We define the conjugate momentum to each mode as the left derivative
\bea
\pi_n &=& L \overleftarrow{\frac{\partial}{\partial \dot\psi_n}}
\eea
The Hamiltonian is
\bea
H &=& \sum_n \(\mu_n + \frac{1}{2r}\) \(N_n - \frac{1}{2}\)
\eea
If we were to decompose this as 
\bea
H &=& H_0 + \frac{1}{2r}\(R_{12} + R_{34}\) + m \(R_{12} - R_{34}\)
\eea
then we would read off 
\bea
H_0 &=& \sum_n \mu_n \(N_n - \frac{1}{2}\)\cr
\frac{1}{2r}\(R_{12}+R_{34}\) &=& \sum_n \frac{1}{2r} \(N_n-\frac{1}{2}\)\cr
R_{12}-R_{34} &=& 0
\eea 
However, we would not be sure how to understand the zero point contribution to the R-charges using such an approach. The contribution to the single particle index is
\bea
-\sum_{n=0}^{\infty} \(f_n e^{-\beta (n+3)} + f_n e^{-\beta (n+2)}\)
\eea
where the exponents are corresponding to $\mu_n + \frac{1}{2r}$. {Again we are not including the zero point energies in the single-particle index}. The negative energy modes $-n-2$ are treated in the standard fashion by filling up the Dirac sea and then they become positive energy modes for the antiparticles.

Using that $f_{-1} = 0$, we may write the sum in a more neat form as
\bea
-\sum_{n=0}^{\infty} \(f_{n-1} + f_n\) e^{-\beta \(n+2\)}
\eea

\subsection{The selfdual tensor field contribution}
The simplest way to see how to quantize the oscillator modes for the tensor field is from the path integral. After BRST quantization, we get the following partition function of a non-selfdual two-form gauge potential,
\bea
Z_{2-form} &=& \frac{\det\triangle_1}{\det^{\frac{1}{2}}\triangle_2 \det^{\frac{3}{2}}\triangle_0}
\eea
{ We now note that the non-harmonic spectrum of $\triangle_p$ separates into exact and coexact parts. To see this, we use Hodge decomposition of a non-harmonic $p$-form, 
\bea
\omega_p &=& d \omega_{p-1} + d^{\dag} \omega_{p+1}
\eea
We then find that 
\bea
\triangle_p \omega_p &=& dd^{\dag}d\omega_{p-1}+d^{\dag}dd^{\dag}\omega_{p+1}
\eea
Thus we have
\bea
\det\triangle_p &=& \det (dd^{\dag})_p \det (d^{\dag}d)_p\cr
&:=& \det \triangle_p^{ex} \det \triangle_p^{coex}
\eea
Moreover, since $\omega_{p-1} = d\omega_{p-2}+d^{\dag}\eta_p+\omega_{p-1}^{harm}$, we have $d\omega_{p-1} = dd^{\dag}\eta_p$ and hence only the coexact part of $\omega_{p-1}$ contributes. Therefore
\ben
\det \triangle_p^{ex} &=& \det \triangle_{p-1}^{coex}\label{extocoex}
\een
and so we have
\ben
\det \triangle_p &=& \det \triangle_p^{coex} \det \triangle_{p-1}^{coex}\label{coex}
\een
Using this, we can write the partition function in the form
\bea
Z_{2-form} &=& \frac{\det^{\frac{1}{2}}\triangle_1^{coex}}{\det^{\frac{1}{2}}\triangle_2^{coex} \det^{\frac{1}{2}}\triangle_0^{coex}}
\eea
Now we assume the six-manifold is on the form $\mb{R}\times S^5$ where time direction is along $\mb{R}$. Then we can decompose a coexact $p$-form as
\bea
Y_p^{coex} &=& Y_p^{coex}+Y_{p-1}^{coex}\wedge dt\cr
\iota_{\partial_t} Y_p^{coex} &=& 0
\eea
Namely, had $Y_{p-1}$ been exact (or harmonic), then also $Y_{p-1}\wedge dt$ would be exact (or harmonic) as well. Thus we have 
\ben
\det\triangle_p^{coex} &=& \det\(\partial_t^2+\triangle_p^{5d,coex}\) \det\(\partial_t^2+\triangle_{p-1}^{5d,coex}\)\label{dimred}
\een
Using this, we find that many factors cancel and we get
\bea
Z_{2-form} &=& \frac{1}{\det^{\frac{1}{2}}\(\partial_t^2+\triangle_2^{5d,coex}\)}
\eea}
This is of the form that we would get from a Hamiltonian that is given by $B \triangle_2^{coex} B$ where $B$ is a coexact two-form. We can expand coexact two-forms in a basis of of two-form spherical harmonics that form a reducible representation of $SO(6)$. This reducible representation separates into one chiral and one antichiral representation. To get the contribution from the chiral two-form we expand in modes from the chiral representation only. These come with the degeneracies $b_n^+=\frac{1}{4} n (n+1)(n+3)(n+4)$ as given in (\ref{degs}), and with corresponding eigenvalues $n(n+4)+4=(n+2)^2$. Then this will contribute with the term
\bea
 \sum_{n=0}^{\infty} b_n^{+} e^{-\beta \(n+2\)}
\eea
to the Witten index. 

\subsection{The tensor multiplet Witten index}
The Witten index is the sum of the individual contributions weighted with a minus sign for the fermions, 
\bea
\I &=& \sum_{n=0}^{\infty} \(b_n + b_n^+ - f_{n-1} - f_n\) e^{-\beta (n+2)}\cr
&=& - \sum_{n=0}^{\infty} (n+1)(n+3) e^{-\beta (n+2)}\cr
&=& \frac{e^{-3\beta}-3e^{-2\beta}}{\(1-e^{-\beta}\)^3} 
\eea
On the other hand, for the particle partition function there is no minus sign,
\bea
\Z &=& \sum_{n=0}^{\infty} \(b_n + b_n^+ + f_{n-1} + f_n\) e^{-\beta (n+2)}\cr
 &=& \frac{e^{-5\beta} - 5 e^{-4\beta} + 15 e^{-3\beta} + 5 e^{-2\beta}}{\(1-e^{-\beta}\)^5}
\eea

\subsection{The hypermultiplet scalars contribution}
We group the four hypermultiplet scalars in two pairs. We denote the scalars in one pair as $\phi^a$ where $a=1,2$. We have their Lagrangian as
\bea
\L &=& \frac{1}{2} \dot{\phi}^a \dot{\phi}^a + \mu \epsilon^{ab} \dot\phi^a \phi^b - \frac{1}{2}\phi^a \(M^2 + \Delta\) \phi^a
\eea
where
\bea
M^2 &=& \frac{15}{4r^2} -  \frac{m}{r} - m^2\cr
\mu &=& \frac{1}{2r} + m
\eea
We will refer to the parameter $m$ as the hypermultiplet mass. We define one complex scalar 
\bea
\phi &=& \frac{1}{\sqrt{2}} \(\phi^1 + i \phi^2\)
\eea
and the Lagrangian is
\bea
\L &=& \dot{\bar\phi} \dot\phi - i\mu \(\dot{\bar\phi} \phi - \dot\phi \bar\phi\) - \bar\phi\(M^2+\Delta\)\phi
\eea
Conjugate momenta are defined as
\bea
\pi = \frac{\partial \L}{\partial \dot\phi} = \dot{\bar\phi} + i\mu \bar\phi\cr
\bar\pi = \frac{\partial \L}{\partial \dot{\bar\phi}} = \dot\phi - i\mu \phi
\eea
Conserved charge densities are
\bea
\R &=& i \pi \phi - i \bar\pi \bar\phi\cr
\H &=& \(\pi + \frac{\mu}{i} \bar\phi\) \(\bar\pi - \frac{\mu}{i}\phi\) + \bar\phi\(M^2 + \Delta\) \phi
\eea
We expand the complex scalar field in scalar harmonics with complex modes $z_n$,
\bea
\phi &=& \sum z_n Y_n\cr
\bar\phi &=& \sum \bar{z}_n \bar{Y}_n
\eea
The scalar harmonics have the properties 
\bea
\Delta Y_n &=& \lambda_n Y_n\cr
\int_{S^5} Y_n \bar{Y}_{n'} &=& \delta_{nn'}
\eea
where $n$ represents the multi-index $(n,m,m',m'')$ that labels the scalar harmonics. We define $L = \int_{S^5} \L$ and get
\bea
L &=& \sum \(\dot{\bar{z}}_n \dot{z}_n - i \mu \(\dot{\bar{z}}_n z_n - \dot{z}_n \bar{z}_n\) - \omega_n^2 \bar{z}_n z_n\)
\eea
where 
\bea
\omega_n^2 &:=& M^2 + \lambda_n\cr
&=& \frac{1}{r^2}(n+2)^2 - \mu^2
\eea
This type of Lagrangian has been quantized in \cite{Dunne:1989hv}. Conjugate momenta are
\bea
q_n = \frac{\partial L}{\partial \dot{z}_n} = \dot{\bar{z}}_n + i \mu \bar{z}_n\cr
\bar{q}_n = \frac{\partial L}{\partial \dot{\bar{z}}_n} = \dot{z}_n - i \mu z_n
\eea
and by consistent matching of the two ways of computing the conjugate momenta, we get
\bea
\pi &=& \sum q_n \bar{Y}_n\cr
\bar\pi &=& \sum \bar{q}_n Y_n
\eea
Then if we define $H = \int_{S^5} \H$, we get 
\bea
H = H_0 + \mu R_{12}
\eea
where
\bea
H_0 &=& \sum \(q_n \bar{q}_n + \Omega_n^2 \bar{z}_n z_n\)\cr
R_{12} &=& i \sum \(z_n q_n - \bar{z}_n \bar{q}_n\)
\eea
and we define 
\bea
\Omega_n^2 &=& \mu^2 + M^2 + \lambda_n\cr
&=& \frac{1}{r^2} (n+2)^2
\eea
We define oscillators as
\bea
\alpha_n &=& \frac{1}{\sqrt{2\Omega_n}}\(\bar{q}_n + i \Omega_n z_n\)\cr
\beta_n &=& \frac{1}{\sqrt{2\Omega_n}}\(\bar{q}_n - i \Omega_n z_n\)
\eea
and get
\bea
H_0 &=& \sum \Omega_n \(\bar\alpha_n \alpha_n + \bar\beta_n \beta_n\)\cr
R_{12} &=& \sum \(\bar\alpha_n \alpha_n - \bar\beta_n \beta_n\)
\eea
We quantize by choosing some ordering prescription and by imposing the canonical commutation relations 
\bea
[z_n,q_{n'}] &=& i \delta_{nn'}\cr
[\bar{z}_n,\bar{q}_{n'}] &=& i \delta_{nn'}
\eea
Then
\bea
[\bar\alpha_n,\alpha_{n'}] &=& \delta_{nn'}\cr
[\bar\beta_n,\beta_{n'}] &=& \delta_{nn'}
\eea
We now get
\bea
H &=& \sum_n \(\(\Omega_n + \mu\) N_{n,\alpha} + \(\Omega_n - \mu\) N_{n,\beta} + \Omega_n\)\cr
R_{12} &=& \sum_n \(N_{n,\alpha} - N_{n,\beta}\)
\eea
when acting on the state 
\bea
\bigotimes_n \|N_{n,\alpha},N_{n,\beta}\> &=& \bigotimes_n (\bar\alpha_n)^{N_{n,\alpha}} (\bar\beta_n)^{N_{n,\beta}} \|0,0\>_n
\eea
where we have chosen the Weyl ordering for the number operators in the Hamiltonian. More explicitly, the contribution from $\phi^a$ for $a=1,2$ to the Hamiltonian is given by 
\bea
HT &=&  X_n N_{n,\alpha} + Y_n N_{n,\beta} + \frac{n+2}{r} T\cr
X_n &=& \(\frac{n+2}{r}+\mu\)T \cr
Y_n &=& \(\frac{n+2}{r}-\mu\)T 
\eea
We note that the zero point energy is independent of the chemical potentials when we use the Weyl ordering prescription. By also including the contribution from the remaining two scalar field $\phi^3$ and $\phi^4$ in the hypermultiplet for which we shall flip the sign of $m$, we find that the total Hamiltonian can be expressed in the following form
\bea
H &=& H_0 + \frac{1}{2r}\(R_{12}+R_{34}\)+m\(R_{12}-R_{34}\)
\eea
with
\bea
H_0 &=& \sum_n \frac{1}{r}\(n+2\) \(N_{n,\alpha}+N_{n,\beta}+\t N_{n,\alpha}+\t N_{n,\beta}+2\)\cr
R_{12} &=& \sum_n \(N_{n,\alpha}-N_{n,\beta}\)\cr
R_{34} &=& \sum_n \(\t N_{n,\alpha}-\t N_{n,\beta}\)
\eea

\subsection{The hypermultiplet fermion contribution}
The Lagrangian is
\bea
\L &=& -\frac{i}{2} \psi^r \epsilon_{rs} \dot\psi^s + \frac{i}{2} \psi^r \epsilon_{rs} \gamma^m \nabla_m \psi^s + \frac{m}{2} \psi^r \sigma^3_{rs} \psi^s
\eea
where the natural index position is $\psi^s$ for the hyper fermions, where $s=1,2$. That means that $(\sigma^3)^s{}_t$ is the third Pauli matrix, and $\epsilon_{+-} = 1$ is the charge conjugation matrix on the space of $s,t,...$-indices. The Majorana condition is 
\bea
(\psi^{\alpha s})^* &=& \psi^{\beta t} C_{\beta\alpha} \epsilon_{ts}
\eea
We use this to eliminate $\psi^{\alpha -}$ and express the Lagrangian in terms of a single complex valued Dirac spinor $\psi^{\alpha} := \psi^{\alpha +}$ as
\bea
\L &=& - i \psi^* \dot\psi + i \psi^* \gamma^m \nabla_m \psi + m \psi^* \psi
\eea
We now expand the spinor field in a basis of commuting complex spinor harmonics
\bea
\psi^{\alpha} &=& \psi_n(t) \varphi^{\alpha}_n
\eea
where the spinor harmonics satisfy
\bea
-i\gamma^m \nabla_m \varphi_n &=& \mu_n \varphi_n\cr
\int_{S^5} (\varphi_n^{\alpha})^* \varphi_m^{\beta} &=& \delta^{\beta}_{\alpha} \delta_{n,m}
\eea
Inserting this expansion into the Lagrangian, we get
\bea
L &=& \sum_n\(-i\psi^*_n \dot\psi_n - (\mu_n - m) \psi^*_n \psi_n\)
\eea
The conjugate momentum is defined as
\bea
\pi_n := L \overleftarrow{\frac{\partial}{\partial \dot\psi_n}}
\eea
and the Hamiltonian is 
\bea
H &=& \pi \dot\psi - L
\eea
Using Weyl ordering prescription, we get
\ben
H &=& \sum_n (\mu_n - m) \(N_n - \frac{1}{2}\)\label{FH}
\een
which we may express as $H = H_0 + 2m R_{12}$ with
\bea
H_0 &=& \sum_n \mu_n \(N_n - \frac{1}{2}\)\cr
R_{12} &=& \sum_n -\frac{1}{2} \(N_n - \frac{1}{2}\)
\eea
As before for the tensor multiplet fermion, also for the hypermultiplet fermion, on $S^5$, we have the spectrum $\mu_n = \pm \(n+\frac{5}{2}\)$ where the plus sign comes with the degeneracy $f_n = \frac{1}{6} (n+1)(n+2)(n+3)(n+4)$ and the minus sign also comes with the same degeneracy $f_n$. 

\subsection{The hypermultiplet Witten index} 
For the bosonic part we sum over all the states $\otimes_n \|N_{n,\alpha},N_{n,\beta},\t N_{n,\alpha},\t N_{n,\beta}\>$, { subject to the single-particle constraint that the sum of all number operators is one,}
\bea
\I(T,m) &=& \sum_n b_n e^{-i(n+2)\frac{T}{r} \(N_{n,\alpha}+N_{n,\beta}+\t N_{n,\alpha} + \t N_{n,\beta}\)}\cr
&& e^{-\frac{iT}{2r} \(N_{n,\alpha}-N_{n,\beta}+\t N_{n,\alpha} - \t N_{n,\beta}\)}\cr
&& e^{-imT \(N_{n,\alpha} - N_{n,\beta} - \t N_{n,\alpha} + \t N_{n,\beta}\)}
\eea
We notice that the last two factors are given by 
\bea
e^{-\frac{iT}{2r}\(R_{12}+R_{34}\)} e^{-im T\(R_{12}-R_{34}\)}
\eea

We define the bosonic contribution to the single particle index as
\bea
\I_{bosons} &=& \sum b_n e^{-iE(N_{n,\alpha},N_{n,\beta},\t N_{n,\alpha},\t N_{n,\beta}) T}
\eea
where the sum runs over all single particle states, whose corresponding energy levels are given by
\bea
E_n(1,0,0,0) &=& \frac{n+2}{r} + \frac{1}{2r} + m\cr
E_n(0,1,0,0) &=& \frac{n+2}{r} - \frac{1}{2r} - m\cr
E_n(0,0,1,0) &=& \frac{n+2}{r} + \frac{1}{2r} - m\cr
E_n(0,0,0,1) &=& \frac{n+2}{r} - \frac{1}{2r} + m
\eea
respectively. 

Since the hyper fermions have $R_{12} = -R_{34}$, we find a dependence on the hypermultiplet mass. We may express the Hamiltonian as
\bea
H &=& H_0 +  2m R_{12}
\eea
where the factor of $2$ comes from $R_{12}-R_{34} = 2 R_{12}$. However, by simply looking at our explicit expression for the Hamiltonian $H$ in eq (\ref{FH}), we deduce that the fermionic contribution to the single particle index is
\bea
\I_{fermions} &=& - \sum_n f_n \(e^{-iE^+_n T} + e^{-iE^-_n T}\)
\eea
where 
\bea
E_n^+ &=& \frac{1}{r}\(n+\frac{5}{2}\)-m\cr
E_n^- &=& \frac{1}{r}\(n+\frac{5}{2}\)+m
\eea
Here $E_n^-$ comes from the negative eigenvalue $-n-\frac{5}{2}$ of the Dirac operator on a unit five-sphere, in which case we shall swap the interpretation of the vacuum which brings in an overall sign change of $E_n^-$ that has the effect of changing the sign of $m$.

The index is given by
\bea
\I &=& \sum_{n=0}^{\infty} \(b_{n-1} + b_{n} - f_{n-1}\) e^{-\beta \(n + \frac{3}{2}\)} \(e^{\beta m} + e^{-\beta m}\)
\eea
The partition function is given by
\bea
\Z &=& \sum_{n=0}^{\infty} \(b_{n-1} + b_{n} + f_{n-1}\) e^{-\beta \(n + \frac{3}{2}\)} \(e^{\beta m} + e^{-\beta m}\)
\eea
Explicitly we find that 
\bea
b_{n-1} + b_{n} - f_{n-1} &=& \frac{1}{2} (n+1)(n+2)\cr
b_{n-1} + b_n + f_{n-1} &=& \frac{1}{2} (n+1)(n+2) \frac{1}{3}(2n^2 + 6n + 3)
\eea 
and
\bea
\I &=& \frac{e^{-\frac{3\beta}{2}}}{\(1-e^{-\beta}\)^3}  \(e^{\beta m} + e^{-\beta m}\)\cr
\Z &=& \frac{e^{-\frac{3\beta}{2}}}{\(1-e^{-\beta}\)^3} \(e^{\beta m} + e^{-\beta m}\) \frac{1 + 6 e^{-\beta} + e^{-2\beta}}{\(1-e^{-\beta}\)^2}
\eea

\section{Squashing the Hopf fiber}
Following \cite{Kim:2012tr}, we now compute the refined indices
\bea
\I_{singlet} &=& \tr (-1)^F e^{-iHT} e^{\a(j-3R_{12})}\cr
\I_{triplet} &=& \tr (-1)^F e^{-iHT} e^{\a(j+R_{12})}
\eea
where $j = j_1+j_2+j_3$ is the $U(1)_{Hopf}$ generator that translates along the Hopf fiber of $S^5$. For notational simplicity, we put $\alpha = -\beta a$. Inserting this chemical potential amounts to a squashing along the Hopf fiber that leads to a reduction of isometry group as $SO(6) \rightarrow SU(3) \times U(1)_{Hopf}$. As we show in appendix \ref{singtrip} the supercharges that are preserved by these indices carry charge $j$ either $\frac{3}{2}$ or $j=-\frac{1}{2}$. These correspond to $SO(6)$ isometry spin labels $(---)$ and $\{(++-),(+-+),(-++)\}$ of singlet and triplet supercharges respectively. Since the supercharges carry charge $-\frac{1}{2}$ under $R_{12}$, we see that these indices respect the singlet and triplet supersymmetries respectively. 

Upon reducing along time direction, this will amount to a field theory living on a squashed five-sphere for all fields that are neutral under $R_{12}$. So for example the tensor multiplet tensor field and the scalar field will live on a squashed five sphere. For the fermion, which is charged under $R_{12}$ there will be an additional mass term, but otherwise this will again live on the same squashed five sphere. Thus the fermion mass is correlated with the squashing parameter of the geometry. Here we chose to insert $R_{12}$ instead of the symmetric combination $\frac{1}{2}\(R_{12}+R_{34}\)$. However, since $R_{12} = \frac{1}{2}\(R_{12}+R_{34}\)+\frac{1}{2}\(R_{12}-R_{34}\)$, we see that the indices above are related to the symmetric case by a shift of our mass parameter $m$; for the triplet case we shift $m$ into $m-\frac{a}{2}$, and for the singlet case we shift $m$ into $m+\frac{3a}{2}$ to go to the symmetric case. 

We label $SU(3)$ representations by their Dynkin labels $(p,q)$. Using Weyl's dimension formula, we have the dimension of such a representation as
\bea
\dim(p,q) &=& \frac{1}{2}(p+1)(q+1)(p+q+2)
\eea
We then refine the degeneracies by inserting a chemical potential $\a$ for the $U(1)_{Hopf}$ charge, for the boson harmonics
\bea
b_n(\a) &=& \sum_{p=0}^n \dim(p,n-p) e^{\a(2p-n)}\cr
b_n^+(\a) &=& \sum_{p=0}^{n-1} \(\dim(p,n-p-1) e^{\a(2p-n-2)} + \dim(p,n-p) e^{\a(2p-n)} + \dim(p,n-p+1) e^{\a(2p-n+2)}\)
\eea
and for the fermion harmonics
\bea
f^+_n(\a) &=& \sum_{p=0}^n \(\dim(p,n-p) e^{\a\(2p-n-\frac{3}{2}\)} + \dim(p,n-p+1) e^{\a\(2p-n+\frac{1}{2}\)}\)
\eea
We also have 
\bea
f^-_n(\a) &=& \sum_{p=0}^n \(\dim(p,n-p) e^{\a\(2p-n+\frac{3}{2}\)} + \dim(p+1,n-p) e^{\a\(2p-n-\frac{1}{2}\)}\)\cr
\eea
which does not enter our problem. Let us anyway notice that
\bea
f^-_n(\a) &=& f^+(-\a)\cr
b^-_n(\a) &=& b^+(-\a)
\eea
with the obvious definition for $b^-_n(\a)$. We will give the physics interpretation of these relations at the end of the next subsection.

\subsection{Tensor}
For the tensor multiplet, the bosonic fields carry no R charges. However, the fermionic fields do carry some R charge. For the tensor multiplet we have that $R_{12}-R_{34}=0$ since there is no dependence on the hypermass $m$. We can also see this from the Weyl projection
\bea
\h\Gamma^{1234} \psi &=& -\psi\cr
\h\Gamma^{1234} \epsilon &=& -\epsilon
\eea
which means that $\h\Gamma^{12} \psi = \h\Gamma^{34} \psi$ for the vector fermion. However, for the hyper fermion we have $\h\Gamma^{12} \psi = -\h\Gamma^{34} \psi$ so this will pick up dependence on $m$ when we compute the usual standard index. From $R_{12}=R_{34}=\frac{1}{2}\(N_n-\frac{1}{2}\)$ we conclude that for a single particle excitation $N_n=1$ for some $n$, we have $R_{12}=R_{34}=\frac{1}{2}$. { (We use the short-hand notation $R_{12}$ for the single-particle contribution $R_{12}(N_n=1) - R_{12}(N_n=0)$)}. Thus we shall shift $j$ by the amount $j-3R_{12} = j - \frac{3}{2}$ for the fermions. But we also should keep in mind that we swap the sign of the number operator itself in the second term. Thus we have the refined indices 
\ben
\I_{triplet}(\beta,a) &=& \sum_{n=0}^{\infty} \(b_n(a) + b_n^+(a) - f_{n-1}(a)e^{\a/2} - f_n(a)e^{-\a/2}\) e^{-\beta (n+2)}\label{A}
\een
and 
\ben
\I_{singlet}(\beta,a) &=& \sum_{n=0}^{\infty} \(b_n(a) + b_n^+(a) - f_{n-1}(a)e^{-3\a/2} - f_n(a)e^{3\a/2}\) e^{-\beta (n+2)}\label{B}
\een
We find the results
\bea
\I_{triplet}(\beta,a) &=& \frac{e^{-\a}e^{-3\beta}-e^{-2\a}e^{-2\beta}-2e^{-2\beta}}{\(1-e^{-\a}e^{-\beta}\)^2 \(1-e^{\a} e^{-\beta}\)}
\eea
and
\bea
\I_{singlet}(\beta,a) &=& \frac{e^{3\a} e^{-3\beta} - 3 e^{2\a} e^{-2\beta}}{\(1-e^{\a} e^{-\beta}\)^3}
\eea
Although this is not clear from radial quantization, here it looks as if the different structures in the denominators for singlet and triplet cases does reflect the spins $(---)$ and $\{(++-),(+-+),(-++)\}$ of the corresponding supercharge. That would account for the structures $\(1-e^{\a} e^{-\beta}\)^3$ and $\(1-e^{-\a}e^{-\beta}\)^2 \(1-e^{\a} e^{-\beta}\)$ respectively. Also, if we change the 6d chirality, say by replacing $(---)$ by $(+++)$, then the sign of the chemical potential $a$ gets flipped. This is in accordance with the branching rules under $SO(6)\rightarrow SU(3) \times U(1)_{Hopf}$, where the $U(1)_{Hopf}$ charges are flipped for the $SU(4)$ representation $(\Lambda_3,\Lambda_2,\Lambda_1)$ as compared to the $U(1)_{Hopf}$ charges we get under branching of the $SU(4)$ representation $(\Lambda_1,\Lambda_2,\Lambda_3)$. In other words, 6d chirality and $U(1)_{Hopf}$ charges are correlated. Changing $a$ to $-a$ takes 6d $(2,0)$ theory to 6d $(0,2)$ theory.

\subsection{Hyper}
For the hyper we have $R_{12} = - R_{34}$ and from the form of the Hamiltonian for the hyper fermion we deduce that for a one-fermi particle excitation we have $R_{12} = -\frac{1}{2}$. But now also the scalar fields carry R-charges. 

The index is given by $\I=\I_B-\I_F$ where for the singlet case
\bea
\I_{B,singlet}(\beta,a) &=& \sum_{n=0}^{\infty} \[\(b_{n-1}(a)e^{-3\a} + b_{n}(a)\) e^{-\beta m} + \(b_{n}(a) e^{3\a} + b_{n-1}(a)\) e^{\beta m}\] e^{-\beta \(n+\frac{3}{2}\)}\cr
\I_{F,singlet}(\beta,a) &=& \sum_{n=0}^{\infty} f_{n-1}(a) \(e^{\beta m}e^{\frac{3\a}{2}} + e^{-\beta m} e^{-\frac{3\a}{2}}\) e^{-\beta \(n+\frac{3}{2}\)}
\eea
and for the triplet case
\bea
\I_{B,triplet}(\beta,a) &=& \sum_{n=0}^{\infty} \[\(b_{n-1}(a)e^{\a} + b_{n}(a)\) e^{-\beta m} + \(b_{n}(a) e^{-\a} + b_{n-1}(a)\) e^{\beta m}\] e^{-\beta \(n+\frac{3}{2}\)}\cr
\I_{F,triplet}(\beta,a) &=& \sum_{n=0}^{\infty} f_{n-1}(a) \(e^{\beta m}e^{-\frac{\a}{2}} + e^{-\beta m} e^{\frac{\a}{2}}\) e^{-\beta \(n+\frac{3}{2}\)}
\eea

We get
\bea
\I_{singlet} &=& \frac{e^{-\frac{3\beta}{2}}\(e^{-\beta m} + e^{3\a} e^{\beta m}\)}{\(1-e^{\a} e^{-\beta}\)^3}
\eea
and 
\bea
\I_{triplet} &=& \frac{e^{-\frac{3\beta}{2}}\(e^{-\beta m} + e^{-\a} e^{\beta m}\)}{\(1-e^{\a} e^{-\beta}\)\(1-e^{-\a}e^{-\beta}\)^2}
\eea
One may note that by shifting $m$ as we described above, we can get to the symmetric case. By adding the tensor and hyper contributions, it is then easy to see that our indices agree with the indices $\I^{singlet}_L(\beta,a)$ and $\I^{triplet,II}_L(\beta,a)$ that we obtained in the Introduction from radial quantization.

{ We note that the result we get using Hamiltonian quantization is a result from a straightforward computation, up to one subtle point. Namely the choice of chiralities for the spinor harmonics and for the tensor harmonics. We get different answers depending on how we choose these chiralities. We pick one chirality of the spinor. Then we adjust the chirality of the tensor gauge field harmonics so that we get an index like quantity. If we had chosen the opposite chirality two-form harmonics, we would not have seen the huge cancelation between bosons and fermions. There is now only one choice left, and this is the choice of chirality of the spinor harmonics. If we pick the other chirality, this just amounts to flipping the sign of the chemical potential $\alpha$. So in conclusion, our computation gives a non-ambiguous result up to a sign flip of the chemical potential $\alpha$. The only difference between the singlet and the triplet case is the change from $3\alpha$ to $-\alpha$ when relating the expressions in (\ref{A}) and (\ref{B}) for the tensor multiplet (and we have the same type of relation for the hypermultiplet case). This is a very simple change to make in a Mathematica file. By doing this simple change, we map the singlet index into the triplet index. Since we matched our singlet index with the result from radial quantization, we are confident about the correctness of our result for the triplet case as well, despite this does not match with the expected result from radial quantization. We get $\I^{triplet,II}_L$ instead of the expected result $\I^{triplet,I}_L$. In Hamiltonian quantization we consider $\mb{R} \times S^5$ with Lorentzian time along $\mb{R}$. To relate with radial quantization, we need to Wick rotate time. For the tensor multiplet, we make a change associated with the fermions which carry R-charges when relating singlet and triplet cases. Fermions are sensitive to the signature in the sense that they can be real only in Lorentzian signature. Since we need the Wick rotation that makes fermions complex, there is no direct way to relate Hamiltonian quantization with radial quantization.}

\section{The supersymmetric Casimir energy}
The supersymmetric Casimir energy is defined as the sum of zero point energies $\frac{1}{2} E_n$ for the bosons plus the zero point point energies $-\frac{1}{2} E_n$ for the fermions.\footnote{{We are dealing with a free theory and hence we have harmonic oscillators. Here $E_n$ denote the frequencies of these harmonic oscillators. The corresponding zero point energies when using the Weyl ordering prescription, are then given by $\pm \frac{1}{2} E_n$ for bosonic and fermionic harmonic oscillators respectively. {Instead of Weyl ordering one could use some other prescription. However, the total ground state energy can not change due to supersymmetry which implies a lower bound on the ground state energy. We therefore believe that any other ordering prescription will give the same result for any physical quantity we compute in a supersymmetric theory.}}} However this sum is divergent. We regulate can regulate the sum in a supersymmetric way as
\bea
E &=& \lim_{\epsilon\rightarrow 0} \frac{1}{2} \tr (-1)^F H e^{-\epsilon H}\cr
&=& -\frac{1}{2}\lim_{\epsilon\rightarrow 0}  \frac{\partial}{\partial \epsilon} \tr (-1)^F e^{-\epsilon H}
\eea
Our Hamiltonian is given by 
\bea
H &=& H_0 + \frac{1}{2r} \(R_{12}+R_{34}\) + m \(R_{12}-R_{34}\)
\eea
In \cite{Assel:2015nca} it was argued that the supersymmetric Casimir energy is unambiguous. However, as was noted in \cite{Kim:2013nva}, here it appears to ambiguous since we could pick another regulator. For example, it appears we could use a different mass parameter $m'$ in the regulator Hamiltonian. If we pick $m'=m$, then after subtracting divergent terms and taking $\epsilon$ to zero, we obtain the following supersymmetric Casimir energies for the tensor multiplet and the hypermultiplet 
\bea
E_{tensor} &=& -\frac{11}{240} \cr
E_{hyper} &=& -\frac{17}{1920} + \frac{m^2}{16} - \frac{m^4}{24}
\eea
We then notice that at $m=\frac{1}{2}$ we get $E_{hyper} = \frac{1}{240}$ and so we get in total $E_{tensor} + E_{hyper} = -\frac{1}{24}$. The full index then takes the form of the inverse of the Dedekind eta function. This was first observed in \cite{Kim:2012ava}. Here we see that this result is valid for the Lorentzian M5 brane. At $m=\frac{1}{2}$ we should take $m'=m$ in order to keep all supersymmetry. If we pick $m'$ different from $m$ at that point, then we break half the supersymmetry. We will pick $m'=m$ also for a generic values of $m$. This choice of $m'$ can be justified by matching the resulting full superconformal index (including the Casimir energy factor) with the 5d partition function \cite{Kim:2012qf}. 

\subsection{Squashed five-sphere}
We consider the singlet case indices with generic squashing parameters,
\bea
\I_{tensor}(\beta,a_i) &=& \frac{e^{-3\beta}e^{-3\beta a} -e^{-2\beta}e^{-3\beta a}\(e^{\beta a_3}+e^{\beta a_2}+e^{\beta a_1}\)}{\(1-e^{-\beta(1+a_1)}\)\(1-e^{-\beta(1+a_2)}\)\(1-e^{-\beta (1+a_3)}\)}
\eea
and
\bea
\I_{hyper}(\beta,m,a_i) &=& \frac{e^{-\frac{3\beta}{2}}e^{-\frac{3\beta a}{2}}\(e^{\beta m} + e^{-\beta m}\)}{\(1-e^{-\beta(1+a_1)}\)\(1-e^{-\beta(1+a_2)}\)\(1-e^{-\beta (1+a_3)}\)}
\eea
where $a = (a_1+a_2+a_3)/3$. At $a = 0$ we reproduce the total Casimir energy that was obtained in \cite{Kim:2012qf} 
\bea
E_{M5} &=& -\frac{1}{24}\(1+\frac{2 a_1a_2 a_3+ \(1-a_1a_2-a_2a_3-a_3a_1\)\delta+\delta^2}{(1+a_1)(1+a_2)(1+a_3)}\)
\eea
with $\delta := \frac{1}{4}-m^2$. This is the sum of a contribution from the tensor,
\bea
E_{tensor} &=& -\frac{11-\frac{11}{2}\(a_1^2+a_2^2+a_3^2\)+29a_1a_2 a_3-\frac{1}{2}\(a_1^4+a_2^4+a_3^4\)}{240(1+a_1)(1+a_2)(1+a_3)}
\eea
and from the hyper,
\bea
E_{hyper} &=& -\frac{17+\(14-40m^2\)\(a_1^2+a_2^2+a_3^2\)+8a_1a_2a_3+4\(a_1^4+a_2^4+a_3^4\)-120m^2+80m^4}{1920(1+a_1)(1+a_2)(1+a_3)}
\eea

For a generic value of $a$, we find
\bea
E_{M5} &=& \frac{1}{384}\frac{N_{M5}(a_1,a_2,a_3,m)}{(1+a_1)(1+a_2)(1+a_3)}
\eea
which is the sum of a contribution from the tensor,
\bea
E_{tensor} &=& \frac{N_{tensor}(a_1,a_2,a_3)}{720(1+a_1)(1+a_2)(1+a_3)}
\eea
and from the hyper,
\bea
E_{hyper} &=& \frac{N_{hyper}(a_1,a_2,a_3,m)}{5760(1+a_1)(1+a_2)(1+a_3)}
\eea
where
\bea
N_{tensor}(a,b,c)&=&
-33 - 44 (a + b + c) - 11(a^2 + b^2 + c^2) - 55 (ab + bc + ca) \cr
&&- 4 (a^3 + b^3 + c^3)  - 5 (a + b + c) (ab + bc + ca) - 
 75 abc \cr
&&- (a^4 + b^4 + c^4) - 15 (a + b + c) abc + 
 5 (a^2b^2 + b^2c^2 + c^2a^2)
 \eea
 \bea
N_{hyper}(a,b,c,m)&=&
-51 - 68 (a + b + c) - 62 (a^2 + b^2 + c^2) - 40 (ab + bc + ca)\cr
&&-  28 (a^3 + b^3 + c^3) - 20 (a + b + c) (ab + bc + ca) + 60 abc \cr
&&- 7 (a^4 + b^4 + c^4)-10 (a^2b^2 + b^2c^2 + c^2a^2)
 \cr
 &&+ 120m^2(3 +2 (a + b + c)+(a^2 + b^2 + c^2)  )- 240 m^4
\eea
\bea
N_{M5}(a,b,c,m)&=&
-21 - 28 (a + b + c) - 10 (a^2 + b^2 + c^2) - 32(ab + bc + ca)\cr
&&- 
 4 (a^3 + b^3 + c^3) - 4 (a + b + c) (ab + bc + ca) -36abc
 \cr
 &&-  (a^4 + b^4 + c^4) +2 (a^2b^2 + b^2c^2 + c^2a^2) -8 (a + b + c) abc
 \cr
 &&+ 8m^2(3 +2 (a + b + c)+(a^2 + b^2 + c^2)  )- 16 m^4
\eea

Now let us take $a_1=a_2=a_3=a$. Then one finds
\bea
E_{M5} &=& -\frac{21 (1 + a)^4 - 24 (1 + a)^2 m^2 + 16 m^4}{384(1 + a)^3}
\eea
 which is the sum of a contribution from the tensor,
\bea
E_{tensor} &=& -\frac{11}{240}(1+a)
\eea
and from the hyper,
\bea
E_{hyper} &=& -\frac{17 (1 + a)^4 - 120 (1 + a)^2 m^2 + 80 m^4}{1920 (1 + a)^3}
\eea
Finally we further take $m=\frac{1}{2}(1+a)$. Then 
\ben
E_{M5} &=& -\frac{1+a}{24}\label{CasimirEnergy}
\een
which is the sum of a contribution from the tensor,
\bea
E_{tensor} &=& -\frac{11}{240}(1+a)
\eea
and from the hyper,
\bea
E_{hyper} &=& \frac{1 + a}{240}
\eea

\section{Anomaly polynomials and Casimir energies}
The anomaly polynomial for a single M5 brane was first obtained in \cite{Witten:1996hc}. Here we also like to obtain the separation of this anomaly polynomial into its tensor multiplet and hypermultiplet contributions. This was recently obtained in \cite{Ohmori:2014kda}. Let us here summarize the result. Anomaly polynomials are conventionally denoted by $I$. We hope this does not cause any confusion with our indices that we also denote by $I$. { The M5 brane embedded in 11 dimensions has a normal  bundle $N$ with structure group $SO(5)$, and a tangent bundle $T$ with structure group $SO(6)$ if we assume Euclidean signature of the M5 brane. }
 
The anomaly polynomial of a real Dirac fermion is
\bea
I_D &=& \frac{1}{2} \ch(S(N)) \h A(T)
\eea
where
\bea
\h A(T) &=& 1 - \frac{p_1(T)}{24} + \frac{7 p_1(T)^2 - 4 p_2(T)}{5760}
\eea
and
\bea
\ch(S(N)) &=& 4 + \frac{p_1(N)}{2} + \frac{p_1(N)^2}{96} + \frac{p_2(N)}{24}
\eea
Multiplying the factors together and picking out the 8-form, we get
\bea
I_D &=& \frac{7 p_1(T)^2 - 4 p_2(T)}{2880} - \frac{p_1(T)p_1(N)}{96} + \frac{p_1(N)^2}{192} + \frac{p_2(N)}{48}
\eea
Adding the anomaly of the selfdual tensor gauge field
\bea
I_A &=& \frac{1}{5760} \(16 p_1(T)^2 - 112 p_2(T)\)
\eea
we get the M5 brane anomaly polynomial 
\bea
I_{M5} := I_A + I_D = \frac{1}{48} \(\frac{1}{4}\(p_1(T)-p_1(N)\)^2-p_2(T)+p_2(N)\)
\eea
If we denote the Chern roots of the tangent bundle as $\omega_i$, then the first and second Pontryagin classes are \cite{Eguchi:1980jx}
\bea
p_1(T) &=& \sum_i \omega_i^2\cr
p_2(T) &=& \sum_{i<j} \omega_i^2\omega_j^2
\eea
Similarly, if we let $\sigma_1$ and $\sigma_2$ be the Chern roots of the normal bundle, then
\bea
p_1(N) &=& \sigma_1^2+\sigma_2^2\cr
p_2(N) &=& \sigma_1^2\sigma_2^2
\eea
We can now express the M5 brane anomaly polynomial in terms of these Chern roots as
\bea
I_{M5} &=& \frac{1}{48} \[\sigma_1^2\sigma_2^2-\sum_{i<j}\omega_i^2\omega_j^2+\frac{1}{4}\(\sum_i \omega_i^2-\sigma_1^2-\sigma_2^2\)^2\]
\eea

We will now extract the anomaly polynomial for the hyper and the tensor multiplets separately, following \cite{Ohmori:2014kda}. The anomaly polynomial of a Dirac fermion is given by
\bea
I_{D} &=& \frac{7 p_1(T)^2}{2880} + \frac{7 p_2(T)}{360} - \frac{p_1(T) p_1(N)}{96} + \frac{p_1(N)^2}{192}
\eea
{ The structure group $SO(5)$ of the normal bundle $N$ is reduced to $SO(4) = SU(2)_L \times SU(2)_R$ when we reduce from $(2,0)$ to $(1,0)$ supersymmetry. For the normal bundles $L$ and $R$ with structure groups $SU(2)_L$ and $SU(2)_R$ we have the relations}
\bea
e(N) &=& c_2(L) - c_2(R)\cr
p_1(N) &=& -2\(c_2(L)+c_2(R)\)
\eea
where $e(N) = \sigma_1 \sigma_2$ is the Euler class of $N$. We then separate the contribution from a Dirac fermion into its normal bundle Weyl components as
\bea
I_{hyper,F} &=& \frac{7 p_1(T)^2 - 4 p_2(T)}{5760} + \frac{c_2(L) p_1(T)}{48} + \frac{c_2(L)^2}{24}\cr
I_{tensor,F} &=& \frac{7 p_1(T)^2 - 4 p_2(T)}{5760} + \frac{c_2(R) p_1(T)}{48} + \frac{c_2(R)^2}{24}
\eea
To see that $I_{D} = I_{hyper,F} + I_{tensor,F}$, we notice that
\bea
c_2(L)^2 + c_2(R)^2 &=& \frac{p_2(N)}{2} + \frac{p_1(N)^2}{8}
\eea
To the tensor multiplet anomaly we also have an additional contribution coming from the tensor gauge field, which is given by
\bea
I_{tensor,A} &=& \frac{1}{5760} \(16 p_1(T)^2 - 112 p_2(T)\)
\eea

\subsection{Dictionary}
If we use the BPS equation (\ref{BPS}) to eliminate $\Delta$, then the Witten index, for the singlet case, can be expressed as \cite{Bobev:2015kza}
\bea
\I_{singlet} &=& \tr(-1)^F e^{-\beta\(\sum_i \omega_i j_i - \sigma_1 R_1 - \sigma_2 R_2\)}
\eea
where we define
\bea
\omega_i &=& 1+a_i
\eea
and
\bea
\sigma_1 &=& \frac{1}{2}(\omega_1+\omega_2+\omega_3) - m\cr
\sigma_2 &=& \frac{1}{2}(\omega_1+\omega_2+\omega_3)  + m
\eea
The dictionary of \cite{Bobev:2015kza} amounts to replacing the Chern roots of the normal bundle in the anomaly polynomial with $\sigma_1$ and $\sigma_2$, and the Chern roots of the tangent bundle with $\omega_i$. If one does this in the anomaly polynomial, and then divides it with the tangent bundle Euler class, which after this substitution becomes $e(T) = \omega_1\omega_2\omega_3$, then it turns out that the result agrees with the Casimir energy, up to a minus sign. 
 
That one shall divide with the Euler class was motivated in \cite{Bobev:2015kza} as a result of having applied the Berline-Vergne fixed point formula\footnote{In this formula we are ignorant about sign factors. One sign could come from relating the Pfaffian with the square root determinant that determines the Euler characteristic, another sign factor comes from an overall factor of $(-2\pi)^3 = -(2\pi)^3$ in 6d. Determining the sign factor is an interesting problem.},
\bea
\frac{1}{(2\pi)^3}\int_{\mb{R}^6} I &=& \sum_p \frac{I|_p}{e(T)|_p}
\eea
where $p$ labels all fixed points, and the evaluation in the right-hand side is done by picking the zero-form component out of $I$ and $e(T)$ at the fixed point $p$. Note that we integrate over the M5 brane world volume $\mb{R}^6$, and before making $I$ an equivariant characteristic class, this was an 8-form. After making $I$ equivariant, this contains all degree forms, and it is the 6-form piece that we are integrating over $\mb{R}^6$. For a simple derivation of this fixed point formula, see for instance section 2.6 in \cite{Szabo:1996md}. In our case the only fixed point is the origin which is the fixed point of the $U(1)^3$ action generated by the Cartan generators $j_i$. Thus the idea is to promote the anomaly polynomial, which is a characteristic class that is $d$-closed, into an equivariant characteristic class that is no longer $d$-closed, but (equivariantly) $d_V$-closed, where $d_V = d+\iota_V$. Here $\iota_V$ refers to the contraction associated with the vector field $V = \sum_{i=1}^3\omega_i \partial_{\varphi^i}$ where $\varphi^i$ denote the angular coordinates that are corresponding to the three Cartan rotation generators $j_i$ in $\mb{R}^6$. We shall then integrate this equivariantly closed version of the anomaly polynomial over $\mb{R}^6$ using the fixed point formula. In this fixed point formula the equivariant Euler class corresponds to the Jacobian $\sqrt{\det(\partial_M V^N)} = \omega_1\omega_2\omega_3$. That is, if we write $V = \omega_1 (x_1 \partial_2 - x_2 \partial_1) + ...$ (dots representing similar terms corresponding to $\omega_2$ and $\omega_3$), then $\partial_1 V^2 = \omega_1$ and so on. Now if we identify this Jacobian as an equivariant Euler class of the tangent bundle, it is natural to associate $\omega_i$ with Chern roots of the tangent bundle. Since $\sigma_1$ and $\sigma_2$ enter the index in much the same way and these are associated with rotations in the normal bundle to the M5 brane, it gets natural to identify these with Chern roots of the normal bundle. We will now confirm that this dictionary works for all cases that we have checked. 

First, this dictionary gives us 
\bea
c_2(L) &=& -m^2\cr
c_2(R) &=& -\frac{1}{4}(3+a_1+a_2+a_3)^2
\eea
Let us begin with the case with $a_1=a_2=a_3=a$. We then get the following anomaly polynomials,
\bea
I_{hyper,F} &=& \frac{17(1+a)^4-120m^2(1+a)^2+80m^4}{1920}\cr
I_{tensor,F} &=& \frac{19(1+a)^4}{240}\cr
I_{tensor,A} &=& -\frac{(1+a)^4}{30}
\eea
Summing the the tensor fermion and the tensor gauge field contributions, we get 
\bea
I_{tensor} = I_{tensor,F} + I_{tensor,A} = \frac{11(1+a)^4}{240}
\eea
whereas for the hyper, we have just the contribution from the fermions, $I_{hyper} = I_{hyper,F}$. We now see that indeed these anomaly polynomials match with corresponding Casimir energies, after we divide by the equivariant Euler class $e(T)=(1+a)^3$,
\bea
\frac{1}{(1+a)^3}I_{tensor} &=& -E_{tensor}\cr
\frac{1}{(1+a)^3}I_{hyper} &=& -E_{hyper}
\eea
Using Mathematica we have confirmed that these kind of relations  
\bea
\frac{1}{\omega_1\omega_2\omega_3} I_{tensor} &=& -E_{tensor}\cr
\frac{1}{\omega_1\omega_2\omega_3} I_{hyper} &=& -E_{hyper}
\eea
hold for generic squashing parameters $a_1,a_2,a_3$ and hypermultiplet mass parameter $m$, where we allow for a generic value of $a = \frac{1}{3}\(a_1+a_2+a_3\)$. 

In \cite{Bobev:2015kza} this relation was shown to hold, but only for the sum $I_{M5}= I_{tensor}+I_{hyper}$ and only at the point $a=0$. Here we did a generalization of their result, and found that the relation between anomaly polynomial and Casimir energy still holds.

\section{Some projected indices}
By knowing the index with squashing parameters $a=b=c$, we can extract the index on Lens spaces $S^5/{\mb{Z}_N}$ by using the formula
\bea
\frac{1}{N}\sum_{k=0}^{N-1} e^{\frac{2\pi i k n}{N}} &=& \sum_{k\in \mb{Z}} \delta_{n,kN}
\eea
to pick out modes with $U(1)_{Hopf}$ charges which are integer multiples of $N$. We can do that by computing
\bea
\I_{singlet}(\beta,N) &=& \sum_{k=0}^{N-1} \I_{singlet}\(\beta,a=\frac{2\pi k i}{N}\)
\eea
{We note that since the chemical potential $a$ couples to the operator $j_1+j_2+j_3-3R_{12}$ rather than just $j_1+j_2+j_3$, this means that the fields which are neutral under $R_{12}$ live on Lens space without any further modification. Fields which carry R charges will therefore also live on Lens space, but for those fields there will be additional terms in the Lagrangian which under dimensional reduction along time will become additional mass terms on Lens space.}

By using Mathematica, we get for the first few values on $N$ the following results
\bea
I_{singlet}(\beta,2) &=& \frac{2 e^{-6\beta} - 12 e^{-4\beta} - 6 e^{-2\beta}}{\(1-e^{-2\beta}\)^3}\cr
I_{singlet}(\beta,3) &=& \frac{-3e^{-9\beta} - 33 e^{-6\beta} - 24 e^{-3\beta}}{\(1-e^{-3\beta}\)^3}\cr
I_{singlet}(\beta,4) &=& \frac{4e^{-12\beta} - 72 e^{-8\beta} - 60 e^{-4\beta}}{\(1-e^{-4\beta}\)^3}\cr
I_{singlet}(\beta,5) &=& \frac{-5e^{-15\beta} + 135 e^{-10\beta} + 120 e^{-5\beta}}{\(1-e^{-5\beta}\)^3}
\eea

We can also compute the index at $N=\infty$, which amounts to projecting to zero charge, where the charge is what multiplies the chemical potential $\a$. For the scalars, picking the zero charge sector out of $b_n(\a)$, we get nonvanishing contributions only from even $n=2k$ where we get
\bea
b_{2k} = \dim(k,k) = (k+1)^3
\eea
For the tensor gauge field the zero charge sector picks out
\bea
b_{2k}^+ = \dim(k+1,k-2)+\dim(k,k)+\dim(k-1,k+2) = 3k(k+1)(k+2)
\eea
For the spinor the zero charge sector is a bit more involved as the spinor is charged under $R_{12}$. For the triplet case, we have the contribution
\bea
F_{n} &:=& f_{n-1}^+(\a)e^{\frac{\a}{2}}+f_{n}^+(\a)e^{-\frac{\a}{2}}\cr
&=& \sum_{p=0}^{n-1} \(e^{\a(2p-n)}\dim(p,n-p-1)+e^{\a(2p-n+2)}\dim(p,n-p)\)\cr
&& + \sum_{p=0}^n \(e^{\a(2p-n-2)}\dim(p,n-p)+e^{\a(2p-n)}\dim(p,n-p+1)\)
\eea
Now we pick the zero charge sector of this. We immediately see that only even $n$ can give zero charge, so we let $n=2k$. Then the zero charge sector gives the contribution
\bea
F_{2k} &=& \dim (k,k-1)+\dim(k-1,k+1)+\dim(k+1,k-1)+\dim(k,k+1) \cr
&=& \frac{1}{2} (2k+1)(2k+2)(2k+3)
\eea
For the singlet case, we have the contribution
\bea
F_{n} &=& f_{n-1}^+(\a)e^{-\frac{3\a}{2}}+f_{n}^+(\a)e^{\frac{3\a}{2}}\cr
&=& \sum_{p=0}^{n-1} \(e^{\a(2p-n-2)}\dim(p,n-p-1)+e^{\a(2p-n)}\dim(p,n-p)\)\cr
&& + \sum_{p=0}^n \(e^{\a(2p-n)}\dim(p,n-p)+e^{\a(2p-n+2)}\dim(p,n-p+1)\)
\eea
and picking the zero charge contribution from this, for $n=2k$, we get
\bea
F_{2k} &= & \dim(k+1,k-2)+2\dim(k,k)+\dim(k-1,k+2)\cr
&=& (k+1)(4k^2+8k+1)
\eea
Then for the triplet case
\bea
b_{2k}+b_{2k}^+ -F_{2k} &=& -2(1+k)
\eea
and for the singlet case
\bea
b_{2k}+b_{2k}^+ -F_{2k} &=& 0
\eea
We then get
\bea
\I_{triplet}(\beta,\infty) &=& -\frac{2e^{-2\beta}}{\(1-e^{-2\beta}\)^2}
\eea
and \footnote {  Here and below note that this vanishing result for the single particle index implies that the full index is simply given by one.}
\bea
\I_{singlet}(\beta,\infty) &=& 0
\eea
The power $2$ in the exponent reflects that we compute an index on one dimension lower, on $\mb{R}\times \mb{CP}^2$, and yet we keep the same amount of supersymmetry. The dimension in the exponent becomes clear when we compute the partition function instead. We get
\bea
\Z_{triplet}(\beta,\infty) &=& 4\frac{e^{-6\beta} + 10 e^{-4\beta} + e^{-2\beta}}{\(1-e^{-2\beta}\)^4}
\eea
and
\bea
\Z_{singlet}(\beta,\infty) &=& 2 \frac{e^{-6\beta}+22e^{-4\beta}+e^{-2\beta}}{\(1-e^{-2\beta}\)^4}
\eea
It is now clear that the power $4$ instead of $5$ now reflects the dimension of $\mb{CP}^2$ as opposed to $S^5$.

For the hyper, let us again project on zero $U(1)_{Hopf}$ charge for the singlet case. We find that the contributions from bosons and the fermions perfectly agree with each other, and is given by
\bea
\I_{B,singlet}(\beta) = \I_{F,singlet}(\beta) &=& \sum_{k=0}^{\infty}\(\dim(k+1,k-2)+\dim(k,k)\)e^{-\beta\(2k+\frac{3}{2}\)}e^{-\beta m}\cr
&& + \sum_{k=0}^{\infty} \(\dim(k-1,k+2)+\dim(k,k)\)e^{-\beta\(2k+\frac{5}{2}\)}e^{\beta m}
\eea
so the single-particle Witten index is zero also for the hypermultiplet in the singlet case. 

\section{Outlook}
It would be interesting to see whether the anomaly polynomial on Lens spaces also matches with corresponding Casimir energies on Lens spaces. Here we leave this as an open problem. However, we can immediately check one special case, namely $\mb{R}\times (S^5/\mb{Z}_{\infty}) = \mb{R}\times \mb{CP}^2$. Here since we found that the single-particle Witten index for the singlet case is zero, the Casimir energy is also zero. This is consistent with the fact that the anomaly polynomial is zero on odd dimensional spaces, and on $\mb{R}\times \mb{CP}^2$ in particular. 

\subsection*{Acknowledgment}

This work was supported in part by NRF Grant 2014R1A1A2053737.

\newpage
\appendix

\section{Representation theory}
In this appendix we use notations and conventions from the Lie algebra book \cite{Cahn}.

\subsection{Representations of $SU(3)$}
The Cartan matrix is 
\bea
A_{ij} &=& \(\begin{array}{cc}
2 & -1\\
-1 & 2
\end{array}\)
\eea
Simple roots have Dynkin labels $\alpha_1 = (2,-1)$ and $\alpha_2 = (-1,2)$. Positive roots are $\alpha_1,\alpha_2,\alpha_1+\alpha_2$ respectively. The sum of all positive roots divided by two, is given by 
\bea
\delta &=& \alpha_1+\alpha_2
\eea
Weyl's dimension formula 
\bea
\dim R_{\Lambda} &=& \frac{\prod_{\alpha} \<\alpha,\Lambda+\delta\>}{\prod_{\alpha} \<\alpha,\rho\>}
\eea
with the product being over all positive roots, gives the dimension of the $SU(3)$ representation with the highest weight $\Lambda = (\Lambda_1,\Lambda_2)$ in Dynkin label notation, as
\bea
\dim R_{(\Lambda_1,\Lambda_2)} &=& \frac{1}{2} \(\Lambda_1+1\)\(\Lambda_2+1\)\(\Lambda_1+\Lambda_2+2\)
\eea

\subsection{Representations of $SU(4)\simeq SO(6)$}
The Cartan matrix for $SU(4)$ is 
\bea
A_{ij} &=& \(\begin{array}{ccc}
2 & -1 & 0\\
-1 & 2 & -1\\
0 & -1 & 2
\end{array}\)
\eea
The isomorphism between the Lie algebras of $SU(4)$ and $SO(6)$ amounts to permuting Dynkin label indices $1$ and $2$. Permuting the first and second rows and then permuting the first and second columns, directly gives us the Cartan matrix of $SO(6)$,
\bea
A_{ij} &=& \(\begin{array}{ccc}
2 & -1 & -1\\
-1 & 2 & 0\\
-1 & 0 & 2
\end{array}\)
\eea
Because of this isomorphism, we can just as well work directly with $SU(4)$ instead of $SO(6)$ representations. Their respective Dynkin labels being related by a permutation of first and second entries. In $SU(4)$ notation, simple roots are $\alpha_1 = (2,-1,-1)$, $\alpha_2 = (-1,2,0)$ and $\alpha_3=(-1,2,0)$. The metric is $g_{ij} = A_{ij}$; the inverse is 
\bea
g^{ij} &=& \frac{1}{4}\(\begin{array}{ccc}
3 & 2 & 1\\
2 & 4 & 2\\
1 & 2 & 3
\end{array}\)
\eea
Positive roots are $\alpha_1,\alpha_2,\alpha_3,\alpha_1+\alpha_2,\alpha_2+\alpha_3,\alpha_1+\alpha_2+\alpha_3$. The sum of positive roots is 
\bea
2\delta &=& 3\alpha_1 + 4 \alpha_2 + 3 \alpha_3
\eea
Weyl's dimension formula gives 
\bea
\dim R_{(\Lambda_1,\Lambda_2,\Lambda_3)} &=& \frac{1}{12} \(\Lambda_1+1\)\(\Lambda_2+1\)\(\Lambda_3+1\)\(\Lambda_1+\Lambda_2+2\)\(\Lambda_2+\Lambda_3\)\(\Lambda_1+\Lambda_2+\Lambda_3+3\)
\eea
from which we find
\ben
&&b_n := \dim R_{(0,n,0)} = \frac{1}{12} (n+1)(n+2)^2 (n+3)\cr
&&f_n^+ := \dim R_{(0,n,1)} = \frac{1}{6} (n+1)(n+2)(n+3)(n+4)\cr
&&v_n := \dim R_{(1,n-1,1)} = \frac{1}{3} n(n+2)^2(n+4)\cr
&&b_n^+ := \dim R_{(0,n-1,2)} = \frac{1}{4} n(n+1)(n+3)(n+4)\label{degs}
\een
For $n=1$ these dimensions have the following interpretations in terms of $SO(6)$ objects,
\bea
&&\dim R_{(0,1,0)} = 6 = {\mbox{vector of $SO(6)$}}\cr
&&\dim R_{(0,0,1)} = 4 = {\mbox{Weyl spinor of $SO(6)$}}\cr
&&\dim R_{(1,0,0)} = 4 = {\mbox{antiWeyl spinor of $SO(6)$}}
\eea  
We notice that $f_n^+$ and $b_n^+$ are chiral (with corresponding anti-chiral representations being $(1,n,0)$ and $(2,n-1,0)$). These chiral representations correspond to the 6d Weyl spinor and the selfdual two-form of the 6d $(2,0)$ theory.

The Casimir operator in the irreducible representation with highest weight $\Lambda$ is given by the formula 
\bea
C_{{\Lambda}} &=& \<\Lambda,\Lambda\> + 2 \<\Lambda,\delta\>
\eea
Up to an overall constant that we drop, we get 
\bea
C_{(0,n,0)} &=& n(n+4)\cr
C_{(1,n-1,1)} &=& n(n+4)+3\cr
C_{(0,n-1,2)} &=& n(n+4)+4
\eea

\subsection{Branching rules}
Here we obtain some branching rules under $SO(6)\simeq SU(4) \rightarrow SU(3) \times U(1)_{Hopf}$.

\subsubsection{Scalar harmonics}
Scalar harmonics are functions on $S^5$ of the form
\bea
Y_n &=& C^{i_1\cdots i_n} x^{i_1} \cdots x^{i_n}
\eea
where $C^{i_1\cdots i_n}$ are symmetric and traceless. Symmetric is obvious since it contracts $x^i$'s that are commuting coordinates in $\mb{R}^6$. Traceless is because of the constraint $x^i x^i = r^2$ that we have on the surface of $S^5$. These functions 
form the irreducible representation $(0,n,0)$ of $SU(4)$. Under $SU(4) \rightarrow SU(3)$, we have the branching rule
\bea
(0,n,0) &\rightarrow & \bigoplus_{p=0}^n (p,n-p)
\eea
as can be easily seen by expanding out the symmetric traceless tensor in a complex basis replacing six real coordinates $x^i$ with three complex coordinates $z^a$ and their complex conjugates $z_a$. The $U(1)_{Hopf}$ charges for these $SU(3)$ representations are then given by $2p-n$. We will introduce the notation
\bea
b_n &=& \dim R_{(0,n,0)}
\eea
We will also need the refinement 
\bea
b_n(\alpha) &=& \sum_{p=0}^n \dim(p,n-p) e^{\alpha(2p-n)}
\eea
We will also need to introduce the degeneracy at a fixed given $U(1)_{Hopf}$ charge $m = 2p-n$, which is given by \cite{Kim:2012ava}
\bea 
b_{n,m} &=& \dim R_{\(\frac{n+m}{2},\frac{n-m}{2}\)}\cr
&=& \frac{1}{8} \((n+2)^2-m^2\)(n+2)
\eea
Here $m=-n,-n+2,\cdots,n-2,n$. The Laplace operator acting on the scalar harmonics has the eigenvalue as
\ben
\triangle Y_n &=& \frac{1}{r^2}n(n+4) Y_n\label{scalarlaplacian}
\een
Up to an overall constant, this eigenvalue is equal to the value of the Casimir operator in the $SU(4)$ representation $(0,n,0)$ for any $n=0,1,2,...$.

\subsubsection{Vector harmonics}
Vector harmonics form the irrep $(1,n-1,1)$ and arise by decomposing the product representation
\bea
(0,n,0) \otimes (0,1,0) &=& (0,n+1,0) \oplus (0,n-1,0) \oplus (1,n-1,1)
\eea
For $n=1$ the interpretation of this decomposition is as follows,
\bea
Y^{i_1} dx^{i_2} &=&  \frac{1}{2} \(Y^{i_1} dx^{i_2} + Y^{i_2} dx^{i_1} - \frac{1}{2}\delta^{i_1 i_2} Y^i dx^i\)\cr
&+& \frac{1}{4} \delta^{i_1 i_2} Y^i dx^i\cr
&+& \frac{1}{2}\(Y^{i_1} dx^{i_2}-Y^{i_2} dx^{i_1}\)
\eea
corresponding to 
\bea
(0,1,0) \otimes (0,1,0) &=& (0,2,0)\oplus (0,0,0) \oplus (1,0,1)
\eea
For $n=2$ we have
\bea
Y^{i_1 i_2} dx^{i_3} &=& \frac{1}{2} \(Y^{i_1 i_2} dx^{i_3} - Y^{i_3 i_1} dx^{i_2} - Y^{i_2 i_3} dx^{i_1} - \frac{1}{4} \(\delta^{i_2 i_3} Y^{i_1 i} dx^i+\delta^{i_3 i_1} Y^{i_2 i} dx^i+\delta^{i_1 i_2} Y^{i_3 i} dx^i\)\)\cr
&+& \frac{1}{8} \(\delta^{i_2 i_3} Y^{i_1 i} dx^i+\delta^{i_3 i_1} Y^{i_2 i} dx^i+\delta^{i_1 i_2} Y^{i_3 i} dx^i\)\cr
&+& \frac{1}{2} \(Y^{i_1 i_2} dx^{i_3} + Y^{i_3 i_1} dx^{i_2} + Y^{i_2 i_3} dx^{i_1}\)
\eea
corresponding to
\bea
(0,2,0) \otimes (0,1,0) &=& (0,3,0)\oplus (0,1,0) \oplus (1,1,1)
\eea

Similarly for $SU(3)$, we have
\bea
(p,q)_{p-q} \otimes (1,0)_{+1} &=& \[(p+1,q)\oplus (p-1,q+1)\oplus (p,q-1)\]_{p-q+1}\cr
(p,q)_{p-q} \otimes (0,1)_{-1} &=& \[(p-1,q)\oplus (p+1,q-1)\oplus (p,q+1)\]_{p-q-1}
\eea
In the first line we have the total $U(1)_{Hopf}$ charge $j=p-q+1$. In the second line we have the total $U(1)_{Hopf}$ charge $j=p-q-1$. Here any representation with a negative entry is discarded, so for instance the representation $(p-1,q+1)$ with $p=0$ will be absent. As an example, for $(p,q)=(1,0)$ we have the following interpretation,
\bea
Y^{a} dz^b &=& Y^{[a} dz^{b]} + Y^{(a} dz^{b)}
\eea
that corresponds to 
\bea
(1,0) \otimes (1,0) &=& (0,1) \oplus (2,0)
\eea
We will now change notation, and instead of the total $U(1)_{Hopf}$ charge for the irrep $(p,q)$, we indicate the shift away from the naive $U(1)_{Hopf}$ charge $p-q$. Using this notation, we have
\bea
(p,q)_{0} \otimes (1,0)_{0} &=& (p+1,q)_0\oplus (p-1,q+1)_{+3}\oplus (p,q-1)_0\cr
(p,q)_{0} \otimes (0,1)_{0} &=& (p-1,q)_0\oplus (p+1,q-1)_{-3}\oplus (p,q+1)_0
\eea

We use the decomposition
\bea
(0,n,0) \otimes (0,1,0) &=& (0,n+1,0) \oplus (0,n-1,0) \oplus (1,n-1,1)
\eea
to derive the branching rule of $(1,n-1,1)$ by using known branching rules of scalar harmonics. We find that 
\bea
(1,n-1,1) &\rightarrow & \bigoplus_{p=0}^{n-1} \Big[(p,n-p-1)_0\oplus (p+1,n-p-1)_{-3}\oplus (p+1,n-p)_0\oplus (p,n-p)_{+3}\Big]
\eea

\subsubsection{Spinor harmonics}
Weyl spinor of positive chirality has component $\psi^{s_1 s_2 s_3}$ with $8 s_1 s_2 s_3 = 1$. It becomes one $SU(3)$ triplet with $U(1)_{Hopf}$ charge $j = s_1+s_2+s_3 = \frac{1}{2}$, and one singlet with $U(1)_{Hopf}$ charge $j = -\frac{3}{2}$. Thus we have
\bea
(1,0,0) &\rightarrow & (0,0)_{-\frac{3}{2}}\oplus (0,1)_{\frac{3}{2}}\cr
(0,0,1) &\rightarrow & (0,0)_{\frac{3}{2}}\oplus (1,0)_{-\frac{3}{2}}
\eea
For spinor harmonics, we have the branching rules
\bea
(1,n,0) &\rightarrow & \bigoplus_{p=0}^n \[(p,n-p)_{-\frac{3}{2}} \oplus (p,n-p+1)_{\frac{3}{2}}\]\cr
(0,n,1) &\rightarrow & \bigoplus_{p=0}^n \[(p,n-p)_{\frac{3}{2}} \oplus (p+1,n-p)_{-\frac{3}{2}}\]
\eea

\subsubsection{Two-form harmonics}
Two-form harmonics decompose into selfdual parts, $(2,n-1,0) \oplus (0,n-1,2)$. The chiral two-form harmonics can be generated from chiral spinor harmonics $(1,n-1,0)$ as
\bea
(1,n-1,0) \otimes (1,0,0) &=& (2,n-1,0) \oplus (1,n-2,1) \oplus (0,n,0) 
\eea
which is a decomposition into two-form, one-form and scalar harmonics. Then we apply the known branching rules on the left-hand side and extract the branching rule
\bea
(2,n-1,0) &\rightarrow & \bigoplus_{p=0}^{n-1} \[(p,n-p-1)_{-3}\oplus (p,n-p)_0 \oplus (p,n-p+1)_3\]
\eea
which is consisent with $n=1$
\bea
(2,0,0) &\rightarrow & (0,0)_{-3}\oplus (0,1)_0 \oplus (0,2)_3
\eea
whose interpretation is 
\bea
[x^i dx^j dx^k]_+ &=& \epsilon^{abc} z_a dz_b dz_b \oplus z_a dz^a dz_b \oplus  \epsilon^{ab(c} z_a dz_b  dz^{d)}
\eea
where $[\cdots]_+$ refers to selfdual part with respect to three indices $ijk$.

\subsection{Consistency checks}
As a consistency check of our charge assignments, let us consider
\bea
(1,0,0) \otimes (1,0,0) &=& (2,0,0) \oplus (0,1,0)
\eea
This reduces to 
\bea
&&\((0,0)_{-\frac{3}{2}}\oplus (0,1)_{\frac{3}{2}}\)\otimes \((0,0)_{-\frac{3}{2}}\oplus (0,1)_{\frac{3}{2}}\)\cr
 &=& (0,0)_{-3}\oplus (0,1)_0 \oplus (0,2)_{+3}\cr
&&\oplus (0,1)_0 \oplus (1,0)_0
\eea
Let us also consider
\bea
(1,1,0)\otimes (1,0,0) &=& (2,1,0)\oplus (1,0,1)\oplus (0,2,0)
\eea
We find that this is consistent with our branching rules above,
\bea
(1,1,0) &\rightarrow &(0,1)_{-\frac{3}{2}} \oplus (1,0)_{-\frac{3}{2}} \oplus (0,2)_{\frac{3}{2}} \oplus (1,1)_{\frac{3}{2}}\cr
(1,0,0) &\rightarrow & (0,0)_{-\frac{3}{2}}\oplus (0,1)_{\frac{3}{2}}\cr
(1,0,1) &\rightarrow & (0,0)_0 \oplus (1,0)_{-3}\oplus (1,1)_0\oplus (0,1)_3\cr
(0,2,0) &\rightarrow & (2,0)_0 \oplus (1,1)_0 \oplus (0,2)_0\cr
(2,1,0) &\rightarrow & (0,1)_{-3}\oplus (1,0)_{-3}\oplus (0,2)_0\oplus (1,1)_0\oplus (0,3)_3\oplus (1,2)_3
\eea

\section{From 6d conformal Killing spinor to 5d Killing spinor}
We start with the 6d conformal Killing spinor equation 
\bea
\nabla_M \epsilon &=& \Gamma_M \eta
\eea
on $\mb{R} \times S^5$. We decompose $M = (0,m)$ where $x^m$ is on $S^5$ and assume the metric is Lorentzian
\bea
ds^2 &=& -dt^2 + d\Omega_5
\eea
We then get
\bea
\nabla_m \epsilon &=& \Gamma_m \Gamma^0 \partial_0 \epsilon
\eea
The integrability condition yields the solution
\bea
\epsilon &=& e^{\frac{i}{2r}t} \E + e^{-\frac{i}{2r}t} \F
\eea
where
\bea
\nabla_m \E &=& \frac{i}{2r} \gamma_m \E\cr
\nabla_m \F &=& -\frac{i}{2r} \gamma_m \F\cr
\eea
where we used
\bea
\Gamma_m &=& \sigma^1 \otimes \gamma_m \otimes 1\cr
\Gamma^0 &=& i \sigma^2 \otimes 1 \otimes 1
\eea
and the 6d Weyl projection $\Gamma \epsilon = -\epsilon$.

\section{From 5d Killing spinor to charged 4d covariantly constant spinor}\label{singtrip}
We decompose $m=(\mu,y)$ where $\mu$ is on base and $y$ is fiber coordinate, and write 
\bea
\gamma_{\mu} &=& \t\gamma_{\mu}+rV_{\mu}\gamma\cr
\gamma_y &=& r\gamma
\eea
where tilde is used for tensors on the base, $V_{\mu}$ is the graviphoton for the fibration. Then the 5d Killing spinor equation splits into 
\bea
\partial_y \E - \frac{r^2}{8} W_{\mu\nu} \gamma^{\mu\nu} \E &=& \frac{i}{2}\gamma \E\cr
\t\nabla_{\mu} \E - \frac{r^2}{8} V_{\mu} W_{\nu\lambda} \gamma^{\nu\lambda} \E + \frac{r}{4} W_{\mu\nu}\gamma^{\nu}\gamma \E &=& \frac{i}{2r} (\t\gamma_{\mu} + r V_{\mu} \gamma) \E
\eea
On $\mb{CP}^2$ we have $W_{12} = W_{34} = \frac{2}{r^2}$. The first equation then becomes
\bea
\partial_y \E &=& \frac{1}{2} \(\gamma^{12}+\gamma^{34}\) \E + \frac{i}{2}\gamma \E
\eea
Let us write
\bea
\gamma^{12} \E^{s_1 s_2} &=& -2is_1 \E^{s_1 s_2}\cr
\gamma^{34} \E^{s_1 s_2} &=& -2is_2 \E^{s_1 s_2}
\eea
Then 
\bea
\partial_y \E &=& -i(s_1+s_2+2s_1s_2) \E
\eea
Let us now return to the second equation and the terms 
\bea
\frac{r}{4} W_{\mu\nu} \t\gamma^{\nu} \gamma \E - \frac{i}{2r} \t\gamma_{\mu} \E
\eea
Let us pick $\mu=1$. Then this becomes
\bea
\frac{1}{2r} \(\t\gamma^2\gamma\E - i \t\gamma^1\E\)
\eea
which vanishes if we pick $s_2=\frac{1}{2}$. Similarly for $\mu=3$ we find the corresponding term vanishes for $s_1=\frac{1}{2}$. Thus by choosing these values, $s_1=s_2=\frac{1}{2}$, we get 
\bea
\t\nabla_{\mu} \E + \frac{3i}{2} V_{\mu} \E &=& 0\cr
\partial_y \E &=& -\frac{3i}{2} \E
\eea
That is, we get a covariantly constant and electrically charged Killing spinor on $\mb{CP}^2$ with electric charge $e=\frac{3}{2}$.

If we study $\F$ instead, then we get
\bea
\partial_y \F &=& -i(s_1+s_2-2s_1s_2) \F
\eea
and we consider now the quantity
\bea
\frac{r}{4} W_{\mu\nu} \t\gamma^{\nu} \gamma \E + \frac{i}{2r} \t\gamma_{\mu} \E
\eea
which vanishes for $s_1=s_2 = -\frac{1}{2}$. We then get 
\bea
\t\nabla_{\mu} \F - \frac{3i}{2} V_{\mu} \F &=& 0\cr
\partial_y \F &=& \frac{3i}{2} \F
\eea

The 6d solution is now
\bea
\epsilon &=& e^{\frac{it}{2r}-\frac{3iy}{2}}\E + e^{-\frac{it}{2r}+\frac{3iy}{2}}\F
\eea

\section{Gauging the R-symmetry}
Let us define a new supersymmetry parameter by
\bea
\epsilon_R &=& g \epsilon
\eea
such that 
\bea
\partial_t \epsilon_R &=& 0\cr
\partial_y \epsilon_R &=& 0
\eea
The original derivative becomes $\nabla_M \epsilon = \nabla_M(g^{-1}\epsilon_R)=g^{-1}\(\nabla_M\epsilon_R+(g\nabla_M g^{-1}) \epsilon_R\)$. We define the covariant derivative
\bea
D_M \epsilon_R &=& \nabla_M \epsilon_R - i A_M \epsilon_R,\cr
A_M &=& i g \nabla_M g^{-1}
\eea
and we have the relation
\bea
\nabla_M \epsilon &=& g^{-1} D_M \epsilon_R
\eea
We have the supersymmetry variations
\bea
\delta \phi^A &=& i \bar\epsilon \Gamma^A \psi\cr
\delta B_{MN} &=& i \bar\epsilon \Gamma_{MN} \psi\cr
\delta \psi &=& \frac{1}{12} \Gamma^{MNP} \epsilon H_{MNP} + \Gamma^M \Gamma_A \epsilon \partial_M \phi^A - \frac{2}{3} \Gamma_A\Gamma^M \nabla_M \epsilon \phi^A
\eea
where 
\bea
\bar\epsilon &=& \epsilon^{\dag} \Gamma^0
\eea
Let us now write these in terms of $\epsilon_R$. First we have 
\bea
\bar\epsilon &=& \bar{\epsilon}_R g
\eea
and then
\bea
\delta \phi^A &=& i \bar\epsilon_R g \Gamma^A \psi\cr
&=& i \bar\epsilon_R (g \Gamma^A g^{-1}) g \psi
\eea
Now we notice the invariance relation
\bea
g\Gamma^A g^{-1} &=& (g^{-1})^{A}{}_B \Gamma^B
\eea
and we get
\bea
\delta \phi^A &=& (g^{-1})^{A}{}_B i \bar\epsilon_R \Gamma^B g \psi
\eea
This motivates us to define new rotated fields
\bea
\phi_R^A &=& g^A{}_B \phi^B\cr
\psi_R &=& g \psi
\eea
In terms of these new fields, we get
\bea
\delta \phi_R^A &=& i \bar\epsilon_R \Gamma^A \psi_R
\eea
We have
\bea
D_M \psi_R &=& \nabla_M \psi_R - i A_M \psi_R
\eea
and we have the relation
\bea
\nabla_M \psi &=& g^{-1} D_M \psi_R
\eea
We have
\bea
\delta \psi &=& \Gamma^M \Gamma_A g^{-1} \epsilon_R \partial_M \phi^A - \frac{2}{3} \Gamma_A \Gamma^M g^{-1} D_M \epsilon_R \phi^A
\eea
and so
\bea
\delta \psi_R &=& \Gamma^M g \Gamma_A g^{-1} \epsilon_R (g^{-1})^A{}_B D_M \phi_R^A - \frac{2}{3} g \Gamma_A g^{-1} \Gamma^M D_M \epsilon_R \phi^A
\eea
We use the invariance of gamma matrices and we get
\bea
\delta \psi_R &=& \Gamma^M \Gamma_A \epsilon_R D_M \phi^A_R - \frac{2}{3} \Gamma_A \Gamma^M D_M \epsilon_R \phi^A_R
\eea
Thus we see that the only change is to replace derivatives by gauge covariant ones.

\end{document}